NUWC-NPT Technical Report 12,490
4 August 2024

# A Normal Variance Mixture Model for Robust Kalman Filtering

Michael J. Walsh
Sensors and Sonar Systems Department

**REVIEW COPY**

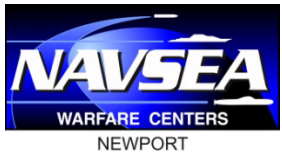

# Naval Undersea Warfare Center Division
# Newport, Rhode Island

# PREFACE

This report was prepared under Naval Undersea Warfare Center Division Newport (NUWCDIVNPT) project, "New Methods for Robust Kalman Filtering," principle investigator Michael J. Walsh (Code 1511). The report was funded by a NUWCDIVNPT Section 219 Basic Research project.

The technical reviewer for this report was Phillip L. Ainsleigh (Code 1511).

Reviewed and Approved:  4 August 2024

Patricia M. Eno
Head, Sensors and Sonar Systems Department

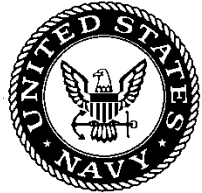

# REPORT DOCUMENTATION PAGE

| 1. REPORT DATE | 2. REPORT TYPE | 3. DATES COVERED | |
|---|---|---|---|
| 04-08-2024 | Technical Report | START DATE | END DATE |

**4. TITLE AND SUBTITLE**

A Normal Variance Mixture Model for Robust Kalman Filtering

| 5a. CONTRACT NUMBER | 5b. GRANT NUMBER | 5c. PROGRAM ELEMENT NUMBER |
|---|---|---|
| | | |
| **5d. PROJECT NUMBER** | **5e. TASK NUMBER** | **5f. WORK UNIT NUMBER** |
| | | |

**6. AUTHOR(S)**

Michael J. Walsh

| 7. PERFORMING ORGANIZATION NAME(S) AND ADDRESS(ES) | 8. PERFORMING ORGANIZATION REPORT NUMBER |
|---|---|
| Naval Undersea Warfare Center Division<br>1176 Howell Street<br>Newport, RI 02841-1708 | TR 12,490 |

| 9. SPONSORING/MONITORING AGENCY NAME(S) AND ADDRESS(ES) | 10. SPONSOR/MONITOR'S ACRONYM(S) | 11. SPONSOR/MONITOR'S REPORT NUMBER(S) |
|---|---|---|
| Naval Undersea Warfare Center Division<br>1176 Howell Street<br>Newport, RI 02841-1708 | NUWCDIVNPT | TR 12,490 |

**12. DISTRIBUTION/AVAILABILITY STATEMENT**

DISTRIBUTION STATEMENT A. Approved for public release; distribution is unlimited.

**13. SUPPLEMENTARY NOTES**

None

**14. ABSTRACT**

The Kalman filter is ubiquitous for state space models because of its desirable statistical properties, ease of implementation, and generally good performance. However, it can perform poorly in the presence of outliers, or measurements with noise variances much greater than those assumed by the filter. An algorithm that is similar to the Kalman filter but robust to outliers is derived in this report. This algorithm—called the normal variance mixture filter (NVMF)—replaces the Gaussian distribution for the noise in the Kalman filter measurement model with a normal variance mixture distribution that admits heavier tails. Choice of the mixing density determines the complexity and performance of the NVMF. When the mixing density is the Dirac delta function, the NVMF is equivalent to the Kalman filter. Choice of an inverse gamma mixing density leads to closed-form recursions for the state estimate and its error covariance matrix that are robust to outliers. The NVMF is compared to the benchmark probabilistic data association filter (PDAF), as well as two other robust filters from the recent literature, for a simulated example. While all four robust filters outperform the Kalman filter when outliers are present, the NVMF provides the most consistent performance across all simulations.

**15. SUBJECT TERMS**

Kalman Filtering, Robust Estimation, Probabilistic Data Association, Outliers, Heavy-Tailed Noise, Normal Variance-Mean Mixture

| 16. SECURITY CLASSIFICATION OF: | | | 17. LIMITATION OF ABSTRACT | 18. NUMBER OF PAGES |
|---|---|---|---|---|
| **a. REPORT**<br>(U) | **b. ABSTRACT**<br>(U) | **C. THIS PAGE**<br>(U) | SAR | 51 |

| 19a. NAME OF RESPONSIBLE PERSON | 19b. PHONE NUMBER (Include area code) |
|---|---|
| Michael J. Walsh | (401) 832-4155 |

**STANDARD FORM 298 (REV. 5/2020)**
*Prescribed by ANSI Std. Z39.18*

(This page has no content.)

# TABLE OF CONTENTS

# TABLE OF CONTENTS (Cont'd)

# LIST OF FIGURES

# LIST OF TABLES

# LIST OF ABBREVIATIONS AND ACRONYMS

| | |
|---|---|
| ANEES | Average normalized estimation error squared |
| CI | Credible interval |
| EM | Expectation-maximization |
| GU | Gaussian-uniform |
| KF | Kalman filter |
| KFOR | Kalman filter for outlier rejection |
| MAP | Maximum *a posteriori* |
| MSE | Mean squared error |
| MMSE | Minimum MSE |
| NRMSE | Normalized root-MSE |
| NVMF | Normal variance mixture filter |
| PCRLB | Posterior Cramér-Rao lower bound |
| PDA | Probabilistic data association |
| PDAF | PDA filter |
| PDF | Probability density function |
| SEM | Supplemented EM |
| VIF | Variational inference filter |

(This page has no content.)

# 1. INTRODUCTION

Kalman filtering is a recursive numerical procedure for estimating the evolving state of a process from sequential measurements of its state; see [1] for Kalman's original derivation of the filter,[*] or any one of the many subsequent references (such as [2]) for an extensive treatment of Kalman filtering theory and filtering theory in general. (The introduction of [2] lists several of the important references to concurrent and independent work by Kalman's contemporaries; see, for example, the work on nonlinear filtering theory by Stratonovich [3].) The process and measurement models for the standard (i.e., linear Gaussian) Kalman filter are given by the following state equations:

$$\mathbf{x}_k = \mathbf{F}_{k-1,k}\mathbf{x}_{k-1} + \mathbf{w}_k, \qquad (1)$$
$$\mathbf{z}_k = \mathbf{H}_k\mathbf{x}_k + \mathbf{v}_k, \qquad (2)$$

where $\mathbf{x}_k \in \mathbb{R}^L$ denotes the state of the process at time index $k$, $\mathbf{F}_{k-1,k}$ is a matrix that describes the evolution of the process from time indices $k-1$ to $k$, $\mathbf{w}_k$ is a process noise term, $\mathbf{z}_k \in \mathbb{R}^M$ is a measurement of the process, $\mathbf{H}_k$ is a matrix that maps the state space to the measurement space, and $\mathbf{v}_k$ is a measurement noise term. The filtering objective is to find the "best" estimate of the state at each update, $k$, given all measurements up through $k$, denoted by $\mathcal{Z}_k = \{\mathbf{z}_1, \ldots, \mathbf{z}_k\}$. This estimate and its error covariance matrix will be denoted by $\hat{\mathbf{x}}_{k|k}$ and $\hat{\mathbf{P}}_{k|k}$, respectively.[**]

The standard filter assumes the process and measurement noise terms are independent and Gaussian distributed with zeros means and known covariance matrices, $\mathbf{Q}_k$ and $\mathbf{R}_k$, respectively. In this report, $p(\cdot)$ will be used to denote a probability density function (PDF) in general, and $\mathcal{N}(\cdot; \boldsymbol{\mu}, \boldsymbol{\Sigma})$ will be used to denote the multivariate Gaussian PDF with mean vector, $\boldsymbol{\mu}$, and covariance matrix, $\boldsymbol{\Sigma}$. Using this notation, the process and measurement noise PDFs are written as

$$p(\mathbf{w}_k) = \mathcal{N}(\mathbf{w}_k; \mathbf{0}, \mathbf{Q}_k), \qquad (3)$$
$$p(\mathbf{v}_k) = \mathcal{N}(\mathbf{v}_k; \mathbf{0}, \mathbf{R}_k). \qquad (4)$$

In practice, the standard filter can perform poorly when these noise assumptions are violated, such as when the actual noise terms are biased or correlated, or when their covariance matrices are different than those assumed, or when the actual noise distributions are non-Gaussian.

In this context of this report, an outlier is a measurement of the state with noise magnitude (variance) much greater than that expected from the Gaussian model assumed by the Kalman filter. An example of a filtering problem with outliers is shown in Figure 1. In this example, the objective is to sequentially estimate the evolving kinematic state of an object moving in one dimension at a constant velocity from periodic measurements of the object's position. The solid blue lines in the figure show the trajectory of the object. The uncircled black dots represent measurements of the object's position, where each measurement is corrupted by a zero mean Gaussian noise with a known, constant variance. The circled dots represent outliers, where each

---

[*]Ironically, Kalman's original formulation assumes perfect measurements (i.e., no measurement noise).
[**]In general, the notation $\hat{\mathbf{x}}_{k|\ell}$ and $\hat{\mathbf{P}}_{k|\ell}$ will be used to denote the estimate of $\mathbf{x}_k$ and its error covariance matrix, respectively, given all measurements up through time index, $\ell$.

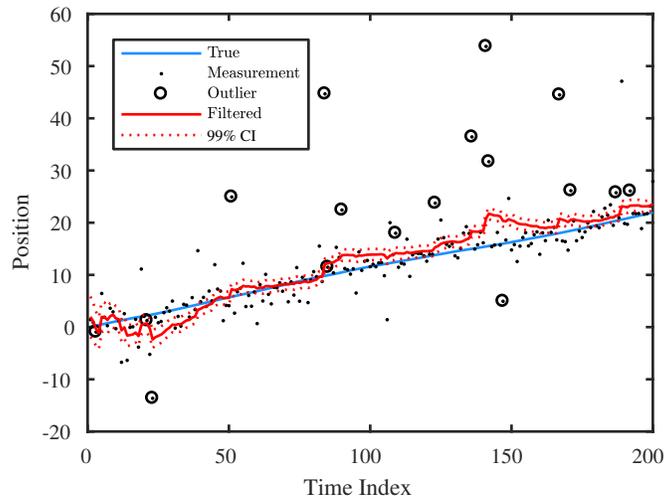

**(a) Kalman Filter**

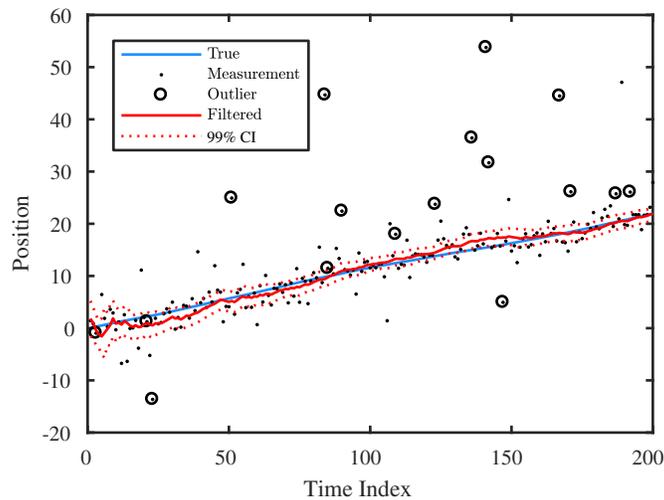

**(b) Robust Filter**

**Figure 1. A One-Dimensional Filtering Problem with Measurement Outliers.**

outlier is corrupted by a zero mean Gaussian noise with an unknown, much larger variance. Application of the standard Kalman filter to this problem results in the estimated trajectory shown by the solid red line in Figure 1(a), with the 99% credible interval (CI) under the Kalman filter statistical assumptions shown by the dotted red lines. It is clear from this example that outliers can have drastic, often abrupt and enduring impacts on the trajectory estimated by the Kalman filter. In contrast, application of the robust filter derived in this report results in the estimated trajectory and CI shown in Figure 1(b). Comparison of the results in Figures 1(a) and 1(b) reveals the robust filter greatly mitigates the impacts of outliers on the estimated trajectory.

The Kalman filter is particularly sensitive to outliers because the Gaussian distribution has relatively light tails. For instance, in the one-dimensional case, only 0.3% of the probability mass of the Gaussian distribution exists beyond three standard deviations to either side of its mean. Consequently, if the measurement, $\mathbf{z}_k$, is an outlier, the standard filter will make an overly large adjustment to the predicted state, $\hat{\mathbf{x}}_{k|k-1}$, so the error between the outlier and the estimated measurement, $\mathbf{H}_k \hat{\mathbf{x}}_{k|k}$, is consistent with the assumed Gaussian statistics of the measurement noise, $\mathbf{v}_k$. These over-adjustments are evident in the example shown in Figure 1(a).

Several authors have developed generalizations of, or modifications to, the Kalman filter to improve its robustness to measurement outliers (and, in some cases, process outliers). These developments, which date from at least as far back as the early 1970s, fall into three general categories. Approaches in the first category maintain the standard Kalman filtering assumptions and framework, but attempt to detect outliers and either discard them, or adapt the filter's assumed noise covariance matrices at each update to better match those of the true noise distributions; see, for example, [4]. Approaches in the second category include an outlier process model along with the model for the process of interest. In these approaches, measurements associate to either the process of interest or the outlier process. This category is exemplified by the well known probabilistic data association (PDA) filter (PDAF) [5]. Approaches in the third category change the distributional assumptions on the measurement (and, in some cases, process) noise(s) to better accommodate outliers (or, as in [6], non-Gaussian noises in general). These approaches include a variety of distributional assumptions for the noise(s), such as:

- discrete Gaussian mixtures for $\mathbf{w}_k$ and $\mathbf{v}_k$ (e.g., [6]),

- Gaussian-uniform mixtures for $\mathbf{v}_k$ (e.g., [7][†]),

- multivariate $t$ distributions for $\mathbf{v}_k$ (e.g., [8–10]), and $\mathbf{w}_k$ and $\mathbf{v}_k$ (e.g., [11, 12]),

- Gaussian scale mixtures for $\mathbf{w}_k$ and $\mathbf{v}_k$ (e.g., [13][‡]).

Each of these approaches requires approximations and, in most cases, iterative procedures to maintain the efficient, recursive structure of the Kalman filter. The approach taken in this report, which falls into the third category, is no exception. It assumes measurement noise has a normal variance mixture distribution (see [14] and Section 2), which admits heavier tails than the Gaussian distribution and, thus, is able to better accommodate outliers (a similar approach is taken in [15] to address outliers in the related Bayesian linear regression problem). Under this assumption, the measurement noise PDF takes the form

$$p(\mathbf{v}_k) = \int_0^\infty \mathcal{N}(\mathbf{v}_k; \mathbf{0}, r\bar{\mathbf{R}}_k) \, p(r) \, dr, \qquad (5)$$

where the positive scalar, $r$, and positive-definite matrix, $\bar{\mathbf{R}}_k$, dictate the scale and shape, respectively, of the covariance matrix of the Gaussian kernel density, and $p(r)$ denotes the mixing density, or prior PDF, of $r$. The filter derived here—called the normal variance mixture filter (NVMF)—treats $r$ as missing data, and uses the expectation-maximization (EM) method [16] and Louis' method [17] to derive recursions for the state estimate, $\hat{\mathbf{x}}_{k|k}$, and its error covariance matrix, $\hat{\mathbf{P}}_{k|k}$, respectively, given the measurement, $\mathbf{z}_k$, and the estimate for the state and its error covariance matrix at time index $k-1$, namely, $\hat{\mathbf{x}}_{k-1|k-1}$ and $\hat{\mathbf{P}}_{k-1|k-1}$ (see Sections 3.1 and 3.2). When the mixing distribution is an inverse gamma distribution (see Equation (17)), these recursions have closed-form solutions (see Section 3.3).

---

[†]This assumption results in a likelihood function similar to that of the PDAF.
[‡]The multivariate $t$ distribution is a special case of the Gaussian scale mixture distribution in [13].

The NVMF shares much in common with those derived in [10] and [13]. The filter in [10] assumes a more general version of the normal variance mixture given by Equation (5), wherein the full covariance matrix of the Gaussian kernel density is treated as a random variable, with (multivariate) mixing density, $p(\mathbf{R}_k)$. Under this assumption, recursions for the state estimate and its error covariance matrix are derived in for both the filtering and smoothing problems for the case when $p(\mathbf{R}_k)$ is an inverse Wishart density, which reduces to the inverse gamma density in the univariate case. While this report focuses on the robust filtering problem, the methods used here may be extended to the robust smoothing problem (see [18]).

The filter in [13] assumes Gaussian scale mixture distributions for both the process and measurement noises and attempts to estimate the parameters of these distributions along with the state at each time index. (The Gaussian scale mixture distribution in [13] is slightly more general than the normal variance-mean mixture distribution in [14], though both admit asymmetry. The normal variance mixture distribution assumed here is a special, symmetric case of the normal variance-mean mixture distribution.) In contrast, the NVMF does not account for process outliers, or measurement noises that follow asymmetric distributions. The tradeoff in flexibility is a much simpler algorithm that is still superior to the standard Kalman filter in the presence of outliers.

Like the NVMF, [10] and [13] and treat the various parameters in their respective mixture models for the noise distributions as "missing data" in the general sense of [16], or more precisely in the contexts of [10] and [13], as latent variables. However, unlike the NVMF, which uses the EM and Louis' methods to derive recursions for the state estimate and its error covariance matrix, [10] and [13] use variational inference to find the approximate posterior density of the latent variables—which include the states—from which these recursions can be derived. In particular, [10] and [13] use variational inference to find the joint density in the "mean-field variational family" of joint densities of the latent variables that minimizes the Kullback-Leibler divergence between this approximation and the true posterior density of latent variables (see [19] for an excellent presentation of variational inference, with a complete example).

As noted in [19], the EM method is a special case of variational inference in the sense that the two methods are equivalent when the joint density of the latent variables is taken as the true posterior density. Variational inference is often more attractive than EM when the true posterior density is hard to compute, and there exists a joint density of the latent variables that closely approximates the posterior and is easier to compute. However, as shown in Section 3.3, the integrals required to compute the posterior density of the missing data for the filtering problem posed in this report are not hard to compute, and a recursion for the state estimate is easily derived using the EM method. Moreover, a recursion for the state estimation error covariance

matrix—requiring little computation beyond that required for the state estimate, and only after the iterative procedure for this estimate has converged—is easily derived using Louis' method.

Lastly, it is explicitly assumed in this report (and, implicitly, in [10] and [13] as well), that the state at time index, $k$, conditioned on all measurements up through the previous time index, $k-1$, is Gaussian distributed with mean given by the predicted state estimate, $\hat{\mathbf{x}}_{k|k-1}$, and covariance matrix given by the predicted state estimation error covariance matrix, $\hat{\mathbf{P}}_{k|k-1}$; that is,

$$p(\mathbf{x}_k|\mathbf{\mathcal{Z}}_{k-1}) = \mathcal{N}(\mathbf{x}_k; \hat{\mathbf{x}}_{k|k-1}, \hat{\mathbf{P}}_{k|k-1}), \tag{6}$$

where

$$\hat{\mathbf{x}}_{k|k-1} = \mathbf{F}_{k-1,k}\hat{\mathbf{x}}_{k-1|k-1}, \tag{7}$$
$$\hat{\mathbf{P}}_{k|k-1} = \mathbf{F}_{k-1,k}\hat{\mathbf{P}}_{k-1|k-1}\mathbf{F}_{k-1,k}^{\mathrm{T}} + \mathbf{Q}_k. \tag{8}$$

While Equation (6) is true under the Gaussian noise models of the Kalman filter, it is not true under the non-Gaussian noise models assumed by the robust filters compared in this report. Nevertheless, this approximation leads to efficient recursions for the state estimate and its error covariance matrix for these robust filters which, in the simulations of Section 4, perform comparably (in most cases) to the Kalman filter in the absence of outliers, and outperform it when outliers are present.

## 2. A HEAVY-TAILED MEASUREMENT NOISE MODEL

In the NVMF, the Gaussian model for measurement noise in the standard Kalman filter (Equation (4)) is replaced with the heavier-tailed, normal variance mixture model (Equation (5)). General properties of the normal variance-mean mixture distribution—of which the normal variance mixture distribution is a special case—are discussed in [14]. For the purpose of this report, it suffices to observe that the measurement noise distribution characterized by Equation (5) is a symmetric distribution with mean zero,

$$\mathrm{E}(\mathbf{v}_k) = \int_{\mathbb{R}^M} \mathbf{v} \left( \int_0^\infty \mathcal{N}(\mathbf{v}; \mathbf{0}, r\bar{\mathbf{R}}_k) \, p(r) \, dr \right) d\mathbf{v}, \qquad (9)$$

$$= \mathbf{0}, \qquad (10)$$

and covariance matrix $\mathrm{E}(r)\bar{\mathbf{R}}_k$,

$$\mathrm{E}(\mathbf{v}_k \mathbf{v}_k^\mathrm{T}) = \int_{\mathbb{R}^M} \mathbf{v}\mathbf{v}^\mathrm{T} \left( \int_0^\infty \mathcal{N}(\mathbf{v}; \mathbf{0}, r\bar{\mathbf{R}}_k) \, p(r) \, dr \right) d\mathbf{v}, \qquad (11)$$

$$= \left( \int_0^\infty r \, p(r) \, dr \right) \bar{\mathbf{R}}_k, \qquad (12)$$

$$= \mathrm{E}(r)\bar{\mathbf{R}}_k. \qquad (13)$$

These results are easily obtained by swapping the order of integration in Equations (9) and (11), and recognizing the inner integrals of the resulting expressions as the first and second moments of the Gaussian distribution, respectively.

Without loss of generality, it will be assumed (as in [14]) that the determinant of the shape matrix, $\bar{\mathbf{R}}_k$, is equal to one. Then the generalized variance associated with the Gaussian kernel density in Equation (5) is $|r\bar{\mathbf{R}}_k| = r^M$. The scale factor, $r$, will often be (loosely) referred to as the measurement noise variance throughout this report. For simplicity, it is assumed that the mixing density, $p(r)$, is the same for each update, $k$, though this assumption is not necessary for the general recursions derived in Sections 3.1 and 3.2.

To illustrate the tail behavior of the normal variance mixture distribution, the PDF in Equation (5) for the one-dimensional case, with $\bar{\mathbf{R}}_k = 1$, is plotted in Figure 2 for the following three choices of the mixing density, $p(r)$:

- Dirac delta function with support at $r_0 > 0$,

$$p(r) = \delta(r - r_0), \quad 0 < r < \infty; \qquad (14)$$

- uniform density with support on $[a, 2r_0 - a]$, with $0 < a < r_0$,

$$p(r) = \begin{cases} \frac{1}{2(r_0-a)}, & a \leq r \leq 2r_0 - a, \\ 0, & \text{otherwise}; \end{cases} \qquad (15)$$

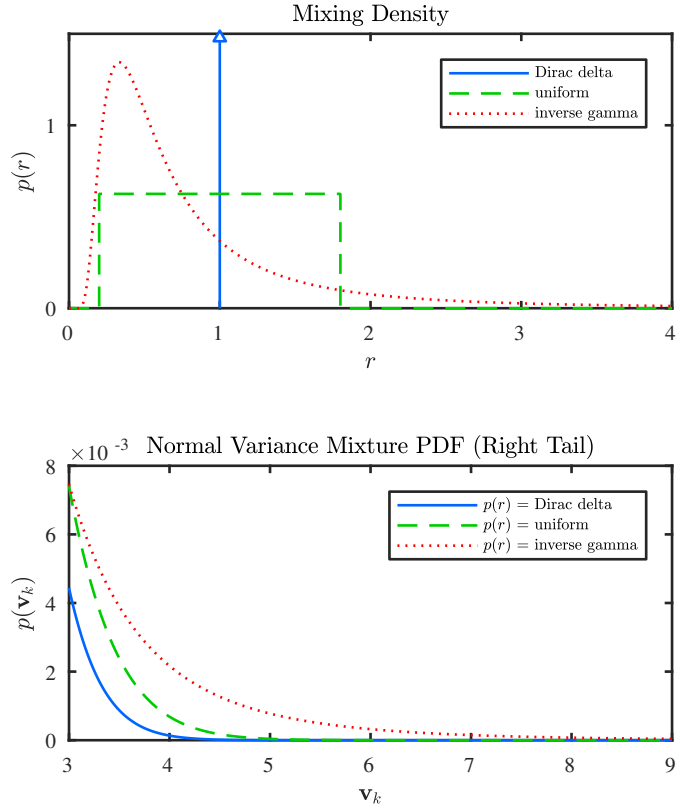

**Figure 2. A One-Dimensional Normal Variance Mixture PDF for Three Mixing Densities.**

- inverse gamma density with shape parameter $\alpha > 1$ and scale parameter $r_0(\alpha - 1) > 0$,

$$p(r) = \frac{[r_0(\alpha - 1)]^\alpha}{\Gamma(\alpha)} \frac{1}{r^{\alpha+1}} e^{-\frac{r_0(\alpha-1)}{r}}, \qquad (16)$$

where $\Gamma$ denotes the gamma function.

Each of these mixing densities has a mean equal to $r_0$. Hence, by Equation (13), the normal variance mixture model of Equation (5) has the same variance for each of these choices of $p(r)$; namely, a variance equal to $r_0$. The top plot in Figure 2 shows the three mixing densities listed above for $r_0 = 1$, $a = 0.2$, and $\alpha = 2$, while the bottom plot shows the right tails of the corresponding normal variance mixture PDFs, each having the same mean (0) and variance (1). When the mixing density is the Dirac delta function, the resulting measurement noise PDF is equivalent to the Gaussian PDF with unit variance. In this case, the NVMF reduces to the standard Kalman filter, which is the appropriate choice if the measurement noise is indeed Gaussian distributed with unit variance. The measurement noise PDFs corresponding to the uniform and inverse gamma mixing densities have much heavier tails than the Gaussian PDF with the same mean and variance; consequently, these noise PDFs are better suited for measurements that may include outliers.

While the choice of the mixing density in the normal variance mixture model for the measurement noise in this problem is, in general, arbitrary, the inverse gamma density is a natural and convenient choice, for reasons discussed below. For an inverse gamma distributed random variable, $r > 0$, with positive shape and scale parameters, $\alpha$ and $\beta$, respectively, the PDF of $r$, denoted by $\mathcal{IG}(r; \alpha, \beta)$, is given by

$$p(r) = \mathcal{IG}(r; \alpha, \beta) = \frac{\beta^\alpha}{\Gamma(\alpha)} \frac{1}{r^{\alpha+1}} e^{-\beta/r}. \tag{17}$$

The inverse distribution for $r$, that is, the distribution for $\tau = 1/r$, is the gamma distribution with positive shape and rate parameters, $\alpha$ and $\beta$, respectively, and PDF, denoted by $\mathcal{G}(\tau; \alpha, \beta)$, given by

$$p(\tau) = \mathcal{G}(\tau; \alpha, \beta) = \frac{\beta^\alpha}{\Gamma(\alpha)} \tau^{\alpha-1} e^{-\beta\tau}. \tag{18}$$

Closed-form expressions for the state estimate and state estimation error covariance matrix recursions for the NVMF make use of the following easily proved identity for positive integer, $n$,

$$\int_0^\infty \frac{1}{r^n} \mathcal{IG}(r; \alpha, \beta)\, dr = \int_0^\infty \tau^n\, \mathcal{G}(\tau; \alpha, \beta)\, d\tau, \tag{19}$$

where the right-hand side is recognized as the $n$th moment of the gamma distributed random variable, $\tau$, with closed-form solution

$$\int_0^\infty \tau^n\, \mathcal{G}(\tau; \alpha, \beta)\, d\tau = \frac{1}{\beta^n} \frac{\Gamma(\alpha+n)}{\Gamma(\alpha)} = \frac{1}{\beta^n} \prod_{i=1}^{n}(\alpha + i - 1), \tag{20}$$

and where the last identity follows from properties of the gamma function.

The inverse gamma density is a natural and convenient choice for the mixing density because of its heavy tails, and because it is the conjugate prior density for $r$; that is, the posterior density of $r$ given all measurements up through time index, $k$, is also inverse gamma (see Section 3.3). The latter property, together with the results of Equations (19) and (20), lead to closed-form expressions for the integrals required by the NVMF. Moreover, as shown in Appendix A, if $r$ is inverse gamma distributed, with shape parameter $\alpha$ and scale parameter $\beta$, then the measurement noise, $\mathbf{v}_k$, under the normal variance mixture model characterized by Equation (5) has the $M$-variate (central) $t$ distribution with degrees of freedom $\nu = 2\alpha$, mean vector $\mathbf{0}$, and correlation matrix $\mathbf{\Sigma} = (\beta/\alpha)\bar{\mathbf{R}}$ (see [20]). Thus, the measurement noise statistics that arise indirectly under the inverse gamma distribution for the measurement noise variance, $r$, are consistent with those of similar approaches that explicitly assume a multivariate $t$ distribution for measurement noise (see [8, 11, 12]).

# 3. THE NORMAL VARIANCE MIXTURE FILTER

## 3.1 STATE ESTIMATE RECURSION

Under the standard Gaussian noise model for measurement noise (Equation (4)), a recursion for the estimate of the state, $\mathbf{x}_k$, given all measurements up through time index $k$, $\boldsymbol{\mathcal{Z}}_k$, is derived in any number or ways; for example, as the minimum mean squared error (MMSE) estimate of $\mathbf{x}_k$ given $\boldsymbol{\mathcal{Z}}_k$, or, equivalently, as the maximum *a posteriori* (MAP) estimate. Under the normal variance mixture model (Equation (5)), a similar recursion is derived as the MAP estimate of $\mathbf{x}_k$ given $\boldsymbol{\mathcal{Z}}_k$ using the EM method. In particular, as shown below, application of the method yields a sequence of Kalman filter-like recursions for the estimate of $\mathbf{x}_k$, for which the corresponding sequence of posterior PDF values is non-decreasing; ideally, the sequence converges to the MAP estimate, though this is not guaranteed. Convergence of the EM algorithm for this problem is discussed in greater detail at the end of this section. A recursion for the estimation error covariance matrix associated with $\hat{\mathbf{x}}_{k|k}$ is derived in Section 3.2.

Under the normal variance mixture model for measurement noise, it is convenient to treat the state estimation problem as an incomplete data problem, where the missing (or, more precisely, unobserved) data at time index, $k$, is the value of the measurement noise variance, $r_k$, treated here as a random variable. The complete data are then the observed measurements, $\boldsymbol{\mathcal{Z}}_k$, and the unobserved value of $r_k$. The complete data posterior PDF for this problem is derived in Appendix B, and is given by Equation (123),

$$p(\mathbf{x}_k|\boldsymbol{\mathcal{Z}}_k, r_k) = \frac{\mathcal{N}(\mathbf{x}_k; \hat{\mathbf{x}}_{k|k-1}, \hat{\mathbf{P}}_{k|k-1})\,\mathcal{N}(\mathbf{z}_k; \mathbf{H}_k\mathbf{x}_k, r_k\bar{\mathbf{R}}_k)}{\mathcal{N}(\mathbf{z}_k; \mathbf{H}_k\hat{\mathbf{x}}_{k|k-1}, r_k\bar{\mathbf{R}}_k + \mathbf{H}_k\hat{\mathbf{P}}_{k|k-1}\mathbf{H}_k^{\mathrm{T}})}. \tag{21}$$

Equation (21) is the posterior PDF for the standard Kalman filter with measurement noise covariance matrix $\mathbf{R}_k = r_k\bar{\mathbf{R}}_k$. Hence, if the value of the measurement noise variance, $r_k$, was truly observed, the recursion for the state estimate, $\hat{\mathbf{x}}_{k|k}$, would be the standard Kalman filter recursion.

The true (incomplete data) posterior PDF for this problem is also derived in Appendix B, and is given by Equation (124),

$$p(\mathbf{x}_k|\boldsymbol{\mathcal{Z}}_k) = \frac{\mathcal{N}(\mathbf{x}_k; \hat{\mathbf{x}}_{k|k-1}, \hat{\mathbf{P}}_{k|k-1}) \int_0^\infty \mathcal{N}(\mathbf{z}_k; \mathbf{H}_k\mathbf{x}_k, r\bar{\mathbf{R}}_k)\, p(r)\, dr}{\int_0^\infty \mathcal{N}(\mathbf{z}_k; \mathbf{H}_k\hat{\mathbf{x}}_{k|k-1}, r\bar{\mathbf{R}}_k + \mathbf{H}_k\hat{\mathbf{P}}_{k|k-1}\mathbf{H}_k^{\mathrm{T}})\, p(r)\, dr}. \tag{22}$$

In general, there is no closed-form solution for the posterior mode of this distribution. The value of $\mathbf{x}_k$ that maximizes Equation (22) may be obtained, in principle, by iterative numerical methods (gradient based or non-gradient based), though the integral in the numerator may make the iterations difficult to compute. Alternatively, the posterior mode of Equation (22) may be obtained using the EM method as follows (see the last paragraph of [16]):

• Expectation step (E-step): Given the estimate of $\mathbf{x}_k$ from the $i$th EM iteration, denoted by $\mathbf{x}_k^{(i)}$, evaluate the conditional expected value of the complete data log-likelihood function,

$$\Upsilon(\mathbf{x}_k|\mathbf{x}_k^{(i)}) = \int_0^\infty [\log p(\mathbf{z}_k, r|\mathbf{x}_k, \boldsymbol{\mathcal{Z}}_{k-1})]\, p(r|\mathbf{x}_k^{(i)}, \boldsymbol{\mathcal{Z}}_k)\, dr, \tag{23}$$

where $p(\mathbf{z}_k, r|\mathbf{x}_k, \mathcal{Z}_{k-1})$, the complete data likelihood function, is derived in Appendix B and is given by Equation (118), and the conditional density, $p(r|\mathbf{x}_k^{(i)}, \mathcal{Z}_k)$, depends on the choice of the mixing density, $p(r)$, for the measurement noise variance, $r$ (this density is derived in Section 3.3 for $p(r)$ equal to the inverse gamma density).

• Maximization step (M-step): Find the estimate of $\mathbf{x}_k$ for the $(i+1)$st iteration such that

$$\mathbf{x}_k^{(i+1)} = \arg\max_{\mathbf{x}_k} \left[ \Upsilon(\mathbf{x}_k|\mathbf{x}_k^{(i)}) + \log p(\mathbf{x}_k|\mathcal{Z}_{k-1}) \right], \tag{24}$$

where the prior density, $p(\mathbf{x}_k|\mathcal{Z}_{k-1})$, is derived in Appendix B and is given by Equation (119).

Starting from an initial estimate for $\mathbf{x}_k$, denoted by $\mathbf{x}_k^{(0)}$, the E- and M-steps are repeated until specified convergence criteria are met (see the end of this section for further discussion of algorithm convergence); ideally, the sequence of EM iterates, $\{\mathbf{x}_k^{(i)}\}$, converges to a stationary point, $\mathbf{x}_k^{(\infty)}$, that maximizes Equation (22), i.e., the MAP estimate.

The EM method is particularly attractive when the M-step has a closed form, as is the case here. Indeed, substituting Equation (118) into Equation (23), taking the logarithm, dropping terms not dependent on $\mathbf{x}_k$, and simplifying the resulting expression yields

$$\Upsilon(\mathbf{x}_k|\mathbf{x}_k^{(i)}) = -\frac{1}{2} (\mathbf{z}_k - \mathbf{H}_k\mathbf{x}_k)^\mathrm{T} \left(\mathbf{R}_k^{(i)}\right)^{-1} (\mathbf{z}_k - \mathbf{H}_k\mathbf{x}_k), \tag{25}$$

where

$$\mathbf{R}_k^{(i)} = \psi(\mathbf{x}_k^{(i)}|\mathcal{Z}_k)\bar{\mathbf{R}}_k, \tag{26}$$

and $\psi$ is a scalar function, given by

$$\psi(\mathbf{x}_k|\mathcal{Z}_k) = \left[\int_0^\infty \frac{1}{r} p(r|\mathbf{x}_k, \mathcal{Z}_k)\, dr\right]^{-1}. \tag{27}$$

Substituting Equations (6) and (25) into Equation (24), taking the logarithm, dropping terms not dependent on $\mathbf{x}_k$, and simplifying the resulting expression yields

$$\mathbf{x}_k^{(i+1)} = \arg\max_{\mathbf{x}_k} \left[ -\frac{1}{2} (\mathbf{z}_k - \mathbf{H}_k\mathbf{x}_k)^\mathrm{T} \left(\mathbf{R}_k^{(i)}\right)^{-1} (\mathbf{z}_k - \mathbf{H}_k\mathbf{x}_k) \right. \\ \left. - \frac{1}{2} (\mathbf{x}_k - \hat{\mathbf{x}}_{k|k-1})^\mathrm{T} \hat{\mathbf{P}}_{k|k-1}^{-1} (\mathbf{x}_k - \hat{\mathbf{x}}_{k|k-1}) \right]. \tag{28}$$

Comparing the term in brackets with the right-hand side of Equation (130) reveals that the M-step for the $(i+1)$st iteration is equivalent to the maximization problem of the MAP formulation of the standard Kalman filter with measurement noise covariance matrix $\mathbf{R}_k = \mathbf{R}_k^{(i)}$. Hence, upon computing $\mathbf{R}_k^{(i)}$ using Equation (26), and given the predicted values $\hat{\mathbf{x}}_{k|k-1}$ and $\hat{\mathbf{P}}_{k|k-1}$, the $(i+1)$st iterate of the state estimate, $\mathbf{x}_k^{(i+1)}$, may be obtained using the standard Kalman filter recursion for the state estimate, namely

$$\mathbf{x}_k^{(i+1)} = \hat{\mathbf{x}}_{k|k-1} + \mathbf{G}_k^{(i)} \left(\mathbf{z}_k - \mathbf{H}_k\hat{\mathbf{x}}_{k|k-1}\right), \tag{29}$$

where

$$\mathbf{G}_k^{(i)} = \hat{\mathbf{P}}_{k|k-1}\mathbf{H}_k^\mathrm{T} \left(\mathbf{R}_k^{(i)} + \mathbf{H}_k\hat{\mathbf{P}}_{k|k-1}\mathbf{H}_k^\mathrm{T}\right)^{-1}, \tag{30}$$

or its information form variant,

$$\mathbf{G}_k^{(i)} = \left[\mathbf{H}_k^\mathrm{T} \left(\mathbf{R}_k^{(i)}\right)^{-1} \mathbf{H}_k + \hat{\mathbf{P}}_{k|k-1}^{-1}\right]^{-1} \mathbf{H}_k^\mathrm{T} \left(\mathbf{R}_k^{(i)}\right)^{-1}, \tag{31}$$

is the Kalman gain matrix from the $i$th iteration. The measurement noise variance estimate at the $i$th iteration, $\psi(\mathbf{x}_k^{(i)}|\boldsymbol{\mathcal{Z}}_k)$, given by Equation (27), depends on the choice of the mixing density for $r$. A closed-form expression for $\psi(\mathbf{x}_k^{(i)}|\boldsymbol{\mathcal{Z}}_k)$ for $p(r)$ equal to the inverse gamma density is given in Section 3.3.

The standard Kalman filter recursion for the estimation error covariance matrix associated with the state estimate at the $(i+1)$st iteration, denoted by $\mathbf{P}_k^{(i+1)}$, and which, in information form, is given by

$$\left(\mathbf{P}_k^{(i+1)}\right)^{-1} = \hat{\mathbf{P}}_{k|k-1}^{-1} + \mathbf{H}_k^\mathrm{T} \left(\mathbf{R}_k^{(i)}\right)^{-1} \mathbf{H}_k, \tag{32}$$

is not required by the M-step. However, upon convergence of the EM algorithm to $\hat{\mathbf{x}}_{k|k} = \mathbf{x}_k^{(\infty)}$, the corresponding estimation error covariance matrix, $\mathbf{P}_{k|k}^{(\infty)}$, forms the basis for the estimation error covariance matrix recursion for $\hat{\mathbf{P}}_{k|k}$, as explained in Section 3.2.

Finally, while the iterates, $\mathbf{x}_k^{(i)}$, are not guaranteed to converge to the MAP estimate, particularly if the posterior PDF in Equation (22) is multi-modal, they do lead to a sequence of posterior PDF values (or, equivalently, log posterior PDF values) that is non-decreasing. Indeed, by extension of Theorem 1 in [16], the log posterior PDF, denoted by $\lambda(\mathbf{x}_k) = \log p(\mathbf{x}_k|\boldsymbol{\mathcal{Z}}_k)$, does not decrease with each iteration; that is, for $i = 0, 1, \ldots,$

$$\lambda(\mathbf{x}_k^{(i+1)}) \geq \lambda(\mathbf{x}_k^{(i)}). \tag{33}$$

Furthermore, if $\{\lambda(\mathbf{x}_k^{(i)})\}$ is a bounded sequence, then it converges monotonically to some finite value, $\lambda^*$. The prior PDF in the numerator of Equation (22) is a Gaussian PDF and, as a function of $\mathbf{x}_k$, is bounded above. For the the mixing density, $p(r)$, equal to the inverse gamma density, the normal variance mixture density in the numerator of Equation (22) is equivalent to the multivariate $t$ density (see Appendix A) and, as a function of $\mathbf{x}_k$, is also bounded above. Thus the log posterior PDF is bounded above for the statistical models assumed in this report, and hence any sequence $\{\lambda(\mathbf{x}_k^{(i)})\}$ from the EM algorithm derived here converges monotonically to a point $\lambda^*$ (which may differ from sequence to sequence, depending on the starting point $\mathbf{x}_k^{(0)}$; see discussion below).

Moreover, by Theorem 2 in [21], if $\Upsilon(\mathbf{x}_k|\mathbf{x}_k^{(i)})$ is continuous in both $\mathbf{x}_k$ and $\mathbf{x}_k^{(i)}$, then $\lambda(\mathbf{x}_k^{(i)})$ converges monotonically to $\lambda^* = \lambda(\mathbf{x}_k^*)$ for some stationary point $\mathbf{x}_k^*$. From Equations (25) through (27), it follows that $\Upsilon(\mathbf{x}_k|\mathbf{x}_k^{(i)})$ is continuous in both $\mathbf{x}_k$ and $\mathbf{x}_k^{(i)}$ provided $\psi(\mathbf{x}_k|\boldsymbol{\mathcal{Z}}_k)$ is continuous. As shown in Section 3.3, for the mixing density, $p(r)$, equal to the inverse gamma density, $\psi(\mathbf{x}_k|\boldsymbol{\mathcal{Z}}_k)$ is indeed continuous (see Equations (55) through (56)). Thus for the statistical models assumed here, the regularity condition of Theorem 2 in [21] holds, and it follows that for the sequence of iterates, $\{\mathbf{x}_k^{(i)}\}$, the corresponding sequence of log posterior PDF values, $\{\lambda(\mathbf{x}_k^{(i)})\}$, converges monotonically to $\lambda^* = \lambda(\mathbf{x}_k^*)$ for some stationary point $\mathbf{x}_k^*$. In general, this stationary point may be the global maximum, a local maximum, or a saddle point of $p(\mathbf{x}_k|\mathcal{Z}_k)$, and the convergence of $\{\lambda(\mathbf{x}_k^{(i)})\}$ to a particular stationary point depends on the choice of the starting point, $\mathbf{x}_k^{(0)}$; in this sense, as noted in [21], the EM algorithm is no different than any other general optimization algorithm, none of which guarantee convergence to the global maximum under general conditions.

Since convergence of the EM algorithm can depend on the choice of starting point, it is often recommended that several EM sequences be generated—each with a different starting point—to determine which stationary point produces the greatest value of $\lambda(\mathbf{x}_k^{(\infty)})$ (see Section 3 of [21]). Alternatively, various modifications and extensions to the EM algorithm that address this issue may be tried (e.g., the deterministic annealing EM algorithm developed in [22]). Despite these recommendations, for this problem it is recommended that a single EM sequence be generated for the state estimate at the current time index, with starting value given by the predicted estimate, namely

$$\mathbf{x}_k^{(0)} = \hat{\mathbf{x}}_{k|k-1}. \qquad (34)$$

This recommendation is largely justified by the effective constraint imposed on $\mathbf{x}_k$ by the Gauss-Markov prior, Equation (6). For the remainder of this report it is assumed that the sequence of EM iterates, $\{\mathbf{x}_k^{(i)}\}$, converges to a stationary point, $\mathbf{x}_k^{(\infty)}$, that maximizes Equation (22), i.e., the MAP estimate.

In practice, the EM iterations may be terminated after increases in the log posterior PDF fall below a specified threshold value, $\epsilon > 0$. From Equation (22), the log posterior PDF is given, in general, by

$$\lambda(\mathbf{x}_k) = -\frac{1}{2}\left(\mathbf{x}_k - \hat{\mathbf{x}}_{k|k-1}\right)^\mathrm{T} \hat{\mathbf{P}}_{k|k-1}^{-1}\left(\mathbf{x}_k - \hat{\mathbf{x}}_{k|k-1}\right) + \log \int_0^\infty \mathcal{N}(\mathbf{z}_k; \mathbf{H}_k\mathbf{x}_k, r\bar{\mathbf{R}}_k)\, p(r)\, dr, \qquad (35)$$

where all terms not dependent on $\mathbf{x}_k$ are dropped, since only differences in the log posterior PDF are of interest here. A closed-form expression for $\lambda(\mathbf{x}_k)$ is given in Section 3.3 for the case when $p(r)$ is equal to the inverse gamma density.

### 3.2 STATE ESTIMATION ERROR COVARIANCE MATRIX RECURSION

Under the standard Gaussian noise model for measurement noise (Equation (4)), a recursion for the state estimation error covariance matrix, $\hat{\mathbf{P}}_{k|k}$, is obtained as a by-product of the derivation for the state estimate, $\hat{\mathbf{x}}_{k|k}$; see, for example, Equations (125) through (129) in Appendix B. This is not the case under the normal variance mixture model (Equation (5)) when the EM method is used to derive a recursion for the state estimate. In this case, obtaining the state estimation error covariance matrix requires extra steps. That the EM method does not produce the estimation error covariance matrix as a by-product is a well known deficiency of the method, and one for which several authors have posed solutions; see, for example, [17] and [23].

In this report, the state estimation error covariance matrix is approximated by the inverse of the observed information matrix [24] (see [25] for a discussion on the merits of using the observed versus expected (Fisher) information for assessing the accuracy of maximum likelihood estimates). The observed information matrix is defined in [24] as the negative second derivative of the log posterior PDF, denoted by

$$\mathbf{J}(\mathbf{x}_k|\mathcal{Z}_k) = -\frac{\partial^2}{\partial \mathbf{x}_k \partial \mathbf{x}_k^\mathrm{T}} \log p(\mathbf{x}_k|\mathcal{Z}_k). \tag{36}$$

The state estimation error covariance matrix is then approximated by the inverse of the observed information matrix evaluated at the MAP estimate, that is,

$$\hat{\mathbf{P}}_{k|k} = \mathbf{J}^{-1}(\hat{\mathbf{x}}_{k|k}|\mathcal{Z}_k) \equiv \mathbf{J}^{-1}(\mathbf{x}_k|\mathcal{Z}_k)\big|_{\mathbf{x}=\hat{\mathbf{x}}_{k|k}}. \tag{37}$$

While it is possible to compute the observed information matrix directly by substituting Equation (35) into Equation (36) and evaluating the necessary derivatives, Louis' method provides an easier approach. To apply this approach, first observe the general expression for the log posterior PDF is obtained from the general expression for the posterior PDF given by Equation (114) in Appendix B:

$$\log p(\mathbf{x}_k|\mathcal{Z}_k) = \log p(\mathbf{x}_k|\mathcal{Z}_{k-1}) + \log p(\mathbf{z}_k|\mathbf{x}_k) - \log p(\mathbf{z}_k|\mathcal{Z}_{k-1}). \tag{38}$$

Let $\mathbf{B}(\mathbf{x}_k|\mathcal{Z}_{k-1})$ and $\mathbf{B}(\mathbf{z}_k|\mathbf{x}_k)$ denote the negative second derivative (curvature) matrices associated with the first two terms in Equation (38), namely, the log prior PDF for $\mathbf{x}_k$ and the log likelihood function for $\mathbf{x}_k$, respectively. Then

$$\mathbf{J}(\mathbf{x}_k|\mathcal{Z}_k) = \mathbf{B}(\mathbf{x}_k|\mathcal{Z}_{k-1}) + \mathbf{B}(\mathbf{z}_k|\mathbf{x}_k), \tag{39}$$

since the last term in Equation (38) is not a function of $\mathbf{x}_k$. Under the Gaussian assumption for the prior distribution of $\mathbf{x}_k$ (see Equation (6)), the first term in Equation (39) is simply the inverse of the predicted state estimation error covariance matrix, that is,

$$\mathbf{B}(\mathbf{x}_k|\mathcal{Z}_{k-1}) = \hat{\mathbf{P}}_{k|k-1}^{-1} \tag{40}$$

(see Appendix C for further details). The second term in Equation (39) can be written in terms of the complete data log likelihood function, $\log p(\mathbf{z}_k, r|\mathbf{x}_k, \mathcal{Z}_{k-1})$, using Louis' method. In particular, let $\mathbf{S}(\mathbf{z}_k, r_k|\mathbf{x}_k)$ and $\mathbf{B}(\mathbf{z}_k, r_k|\mathbf{x}_k)$ denote the gradient vector and curvature matrix, respectively, of the complete data log likelihood function. For an arbitrary function, $f(r_k)$, of the unobserved measurement noise variance, $r_k$, let $\mathrm{E}[f(r_k)|\mathcal{Z}_k]$ denote the conditional expectation,

$$\mathrm{E}\left[f(r_k)|\mathcal{Z}_k\right] = \int_0^\infty f(r)\, p(r|\mathbf{x}_k, \mathcal{Z}_k)\, dr. \tag{41}$$

Then, by equation (3.2) in [17], the second term in Equation (39) is written in terms of complete data statistics as

$$\mathbf{B}(\mathbf{z}_k|\mathbf{x}_k) = \mathrm{E}[\mathbf{B}(\mathbf{z}_k, r_k|\mathbf{x}_k)|\mathcal{Z}_k] - \mathrm{E}\left[\mathbf{S}(\mathbf{z}_k, r_k|\mathbf{x}_k)\mathbf{S}^\mathrm{T}(\mathbf{z}_k, r_k|\mathbf{x}_k)|\mathcal{Z}_k\right] \\ + \mathrm{E}[\mathbf{S}(\mathbf{z}_k, r_k|\mathbf{x}_k)|\mathcal{Z}_k]\, \mathrm{E}\left[\mathbf{S}^\mathrm{T}(\mathbf{z}_k, r_k|\mathbf{x}_k)|\mathcal{Z}_k\right]. \tag{42}$$

Expressions for $\mathbf{S}(\mathbf{z}_k, r_k|\mathbf{x}_k)$ and $\mathbf{B}(\mathbf{z}_k, r_k|\mathbf{x}_k)$ under the normal variance mixture model for measurement noise are derived in Appendix C; see Equations (136) and (137), respectively. Substituting these results into Equation (42), evaluating the expectations, and simplifying the resulting expression yields

$$\mathbf{B}(\mathbf{z}_k|\mathbf{x}_k) = \mathbf{H}_k^{\mathrm{T}} \left[\psi(\mathbf{x}_k|\boldsymbol{\mathcal{Z}}_k)\bar{\mathbf{R}}_k\right]^{-1} \mathbf{H}_k \\ - \mathbf{H}_k^{\mathrm{T}} \left[\phi(\mathbf{x}_k|\boldsymbol{\mathcal{Z}}_k)\bar{\mathbf{R}}_k\right]^{-1} (\mathbf{H}_k\mathbf{x}_k - \mathbf{z}_k)(\mathbf{H}_k\mathbf{x}_k - \mathbf{z}_k)^{\mathrm{T}} \left[\phi(\mathbf{x}_k|\boldsymbol{\mathcal{Z}}_k)\bar{\mathbf{R}}_k\right]^{-1} \mathbf{H}_k, \quad (43)$$

where $\phi$ is a scalar function, given by

$$\phi(\mathbf{x}_k|\boldsymbol{\mathcal{Z}}_k) = \left\{ \int_0^\infty \frac{1}{r^2} p(r|\mathbf{x}_k, \boldsymbol{\mathcal{Z}}_k)\, dr - \left[\int_0^\infty \frac{1}{r} p(r|\mathbf{x}_k, \boldsymbol{\mathcal{Z}}_k)\, dr\right]^2 \right\}^{-1/2}. \quad (44)$$

To simplify notation, let

$$\mathbf{u}_k(\mathbf{x}_k) = \mathbf{H}_k^{\mathrm{T}} \left[\phi(\mathbf{x}_k|\boldsymbol{\mathcal{Z}}_k)\bar{\mathbf{R}}_k\right]^{-1} (\mathbf{H}_k\mathbf{x}_k - \mathbf{z}_k). \quad (45)$$

Then, substituting Equations (40), (43), and (45) into Equation (39), simplifying the result, and evaluating the final expression at the $i$th EM iterate yields

$$\mathbf{J}(\mathbf{x}_k^{(i)}|\boldsymbol{\mathcal{Z}}_k) = \hat{\mathbf{P}}_{k|k-1}^{-1} + \mathbf{H}_k^{\mathrm{T}} \left(\mathbf{R}_k^{(i)}\right)^{-1} \mathbf{H}_k - \mathbf{u}_k(\mathbf{x}_k^{(i)})\mathbf{u}_k^{\mathrm{T}}(\mathbf{x}_k^{(i)}). \quad (46)$$

Hence, using Equation (32), and insofar as $\mathbf{x}_k^{(i)} \to \hat{\mathbf{x}}_{k|k}$ as $i \to \infty$,

$$\hat{\mathbf{P}}_{k|k}^{-1} = \left(\mathbf{P}_k^{(\infty)}\right)^{-1} - \mathbf{u}_k(\hat{\mathbf{x}}_{k|k})\mathbf{u}_k^{\mathrm{T}}(\hat{\mathbf{x}}_{k|k}), \quad (47)$$

or, using the Sherman-Morrison formula,

$$\hat{\mathbf{P}}_{k|k} = \mathbf{P}_k^{(\infty)} + \frac{\mathbf{P}_k^{(\infty)}\mathbf{u}_k(\hat{\mathbf{x}}_{k|k})\mathbf{u}_k^{\mathrm{T}}(\hat{\mathbf{x}}_{k|k})\mathbf{P}_k^{(\infty)}}{1 - \mathbf{u}_k^{\mathrm{T}}(\hat{\mathbf{x}}_{k|k})\mathbf{P}_k^{(\infty)}\mathbf{u}_k(\hat{\mathbf{x}}_{k|k})}. \quad (48)$$

Equations (32) and (48) (or any one of the alternative forms for the latter) constitute the recursion for the state estimation error covariance matrix associated with the state estimate, $\hat{\mathbf{x}}_{k|k}$, for the NVMF. The covariance matrix, $\mathbf{P}_k^{(\infty)}$, is computed from Equation (32) (or, again, any one of its alternative forms) upon convergence of the EM algorithm for $\hat{\mathbf{x}}_{k|k}$. Closed-form expressions for $\psi(\hat{\mathbf{x}}_{k|k}|\boldsymbol{\mathcal{Z}}_k)$ and $\phi(\hat{\mathbf{x}}_{k|k}|\boldsymbol{\mathcal{Z}}_k)$ are given in Section 3.3 for choice of the inverse gamma density for the unobserved measurement noise variance, $r$, in Equation (5).

The covariance matrix, $\mathbf{P}_k^{(\infty)}$, in Equations (47) and (48) is the state estimation error covariance matrix associated with the equivalent Kalman filter for the final M-step of the EM algorithm for $\hat{\mathbf{x}}_{k|k}$. It is clear from Equation (48) that this covariance matrix captures only part of the uncertainty in $\hat{\mathbf{x}}_{k|k}$. In fact, the second term on the right-hand side of Equation (48) accounts for the uncertainty introduced by the fact that the measurement noise variance, $r_k$, is not observed. Indeed, as pointed out by Louis for the general case addressed in [17], Equation (47) is a statement of the missing information principle discussed in [26]. That is, the observed information, $\hat{\mathbf{P}}_{k|k}^{-1}$, is equal to the complete information associated with the EM formulation of the estimation problem, encoded here in the information matrix $(\mathbf{P}_k^{(\infty)})^{-1}$, minus the information introduced by the missing data, encoded in the outer product, $\mathbf{u}_k(\hat{\mathbf{x}}_{k|k})\mathbf{u}_k^{\mathrm{T}}(\hat{\mathbf{x}}_{k|k})$.

Incidentally, application of the alternative approach in [23] to computing the estimation error covariance matrix when using the EM algorithm results in an algebraic form for $\hat{\mathbf{P}}_{k|k}$ that is identical to that of Equation (48) (see equation (2.3.5) in [23]), except their approach, called the Supplemented EM (SEM) algorithm, involves an iterative numerical procedure to calculate the increase in estimation uncertainty due to the missing information, i.e., the second term on the right-hand side of Equation (48). The SEM algorithm is attractive in cases where this term is difficult or impossible to compute analytically. As shown in the next section, for the measurement noise model chosen here, all terms in Equations (47) and (48) are available in closed forms that are easy to compute, and Louis' method is most attractive.

## 3.3 STATEMENT OF FILTER FOR INVERSE GAMMA MIXING DENSITY

The recursions for the state estimate and its error covariance matrix derived in Sections 3.1 and 3.2, respectively, are valid for any choice of the mixing density, $p(r)$, of the normal variance mixture model for measurement noise. In order to explicitly state the steps required to implement a robust filter for this heavy-tailed measurement noise model, the mixing density must be specified.

The recursions for the state estimate and its error covariance matrix (or its inverse), given by Equation (29) and Equation (48) (or Equation (47)), respectively, are functions of the conditional estimates for the measurement noise variance given by Equations (27) and (44), repeated here for convenience:

$$\psi(\mathbf{x}_k|\boldsymbol{\mathcal{Z}}_k) = \left[\int_0^\infty \frac{1}{r} p(r|\mathbf{x}_k, \boldsymbol{\mathcal{Z}}_k)\, dr\right]^{-1}, \tag{49}$$

and

$$\phi(\mathbf{x}_k|\boldsymbol{\mathcal{Z}}_k) = \left\{\int_0^\infty \frac{1}{r^2} p(r|\mathbf{x}_k, \boldsymbol{\mathcal{Z}}_k)\, dr - \left[\int_0^\infty \frac{1}{r} p(r|\mathbf{x}_k, \boldsymbol{\mathcal{Z}}_k)\, dr\right]^2\right\}^{-1/2}. \tag{50}$$

The missing data conditional PDF in these estimates is derived from the general expression for the complete data likelihood function, Equation (104), and is given by

$$p(r|\mathbf{x}_k, \boldsymbol{\mathcal{Z}}_k) = \frac{p(\mathbf{z}_k|\mathbf{x}_k, r)\, p(r)}{\int_0^\infty p(\mathbf{z}_k|\mathbf{x}_k, r)\, p(r)\, dr}. \tag{51}$$

Now, under the normal variance mixture model for the measurement noise in Equation (2), the conditional PDF of $\mathbf{z}_k$ given $\mathbf{x}_k$ and $r$ is given by Equation (117),

$$p(\mathbf{z}_k|\mathbf{x}_k, r) = \mathcal{N}(\mathbf{z}_k; \mathbf{H}_k\mathbf{x}_k, r\bar{\mathbf{R}}_k). \tag{52}$$

Suppose $r$ is inverse gamma distributed with shape parameter, $\alpha$, and scale parameter, $\beta$; that is,

$$p(r) = \mathcal{IG}(r; \alpha, \beta). \tag{53}$$

Substituting Equations (52) and (53) into Equation (51) and manipulating the result, it is straightforward to show

$$p(r|\mathbf{x}_k, \boldsymbol{\mathcal{Z}}_k) = \mathcal{IG}(r; M/2 + \alpha, \zeta(\mathbf{x}_k, \mathbf{z}_k) + \beta), \tag{54}$$

where

$$\zeta(\mathbf{x}_k, \mathbf{z}_k) = \frac{1}{2} (\mathbf{z}_k - \mathbf{H}_k \mathbf{x}_k)^{\mathrm{T}} \bar{\mathbf{R}}_k^{-1} (\mathbf{z}_k - \mathbf{H}_k \mathbf{x}_k). \tag{55}$$

Hence, the inverse gamma distribution is the conjugate prior distribution for $r$. Furthermore, using the identities in Equations (19) and (20), it is straightforward to show that $\psi(\mathbf{x}_k|\boldsymbol{\mathcal{Z}}_k)$ and $\phi(\mathbf{x}_k|\boldsymbol{\mathcal{Z}}_k)$ reduce to the following simple closed forms:

$$\psi(\mathbf{x}_k|\boldsymbol{\mathcal{Z}}_k) = \frac{\zeta(\mathbf{x}_k, \mathbf{z}_k) + \beta}{M/2 + \alpha}, \tag{56}$$

$$\phi(\mathbf{x}_k|\boldsymbol{\mathcal{Z}}_k) = \frac{\zeta(\mathbf{x}_k, \mathbf{z}_k) + \beta}{\sqrt{M/2 + \alpha}}. \tag{57}$$

With these closed forms for $\psi(\mathbf{x}_k|\boldsymbol{\mathcal{Z}}_k)$ and $\phi(\mathbf{x}_k|\boldsymbol{\mathcal{Z}}_k)$, the one-step update for the NVMF is stated simply as an iterative sequence of standard Kalman filters, where each iteration in the sequence uses a refined estimate of the measurement noise covariance matrix. Pseudo-code for the NVMF is provided in Figure 3.

Upon convergence, the state estimation error covariance matrix is computed as a correction to that obtained from the Kalman filter from the terminal EM iteration. The iterations at each time index, $k$, are terminated when increases in the log posterior PDF, $\lambda(\mathbf{x}_k)$, fall below a specified threshold value, $\epsilon > 0$. For choice of the inverse gamma density for the mixing density, the log posterior PDF given by Equation (35) reduces to the closed form

$$\lambda(\mathbf{x}_k) = -\frac{1}{2} (\mathbf{x}_k - \hat{\mathbf{x}}_{k|k-1})^{\mathrm{T}} \hat{\mathbf{P}}_{k|k-1}^{-1} (\mathbf{x}_k - \hat{\mathbf{x}}_{k|k-1})$$
$$- (M/2 + \alpha) \log \left[ 1 + \frac{1}{2\beta} (\mathbf{z}_k - \mathbf{H}_k \mathbf{x}_k)^{\mathrm{T}} \bar{\mathbf{R}}_k^{-1} (\mathbf{z}_k - \mathbf{H}_k \mathbf{x}_k) \right], \tag{58}$$

which is easily computed at each iterate, $\mathbf{x}_k^{(i)}$.

Lastly, as with the standard Kalman filter, it is best practice to implement the robust filter derived here in a numerically stable, "square root" form as discussed extensively in [27]. For example, the square root information form of the Kalman filter reported in [28]—which uses only numerically robust QR decompositions—is easy to implement and a particularly attractive option here, since the state estimation error covariance matrix recursion is most naturally stated in information form (see Equation (47)). Specifically, when the filter is implemented in square root information form, in which the Cholesky factor, $\hat{\mathbf{P}}_{k|k}^{-1/2}$, is propagated instead of $\hat{\mathbf{P}}_{k|k}$, then Equation (47) is preferable since $\hat{\mathbf{P}}_{k|k}^{-1}$ need not be computed explicitly. Instead $\hat{\mathbf{P}}_{k|k}^{-1/2}$ can be computed from the Cholesky factor $(\mathbf{P}_k^{(\infty)})^{-1/2}$ and the vector $\mathbf{u}_k(\hat{\mathbf{x}}_{k|k})$ using a stable and efficient rank-1 Cholesky downdate procedure (see [29]).

1: Set $\mathbf{x}_k^{(0)} = \hat{\mathbf{x}}_{k|k-1} = \mathbf{F}_{k-1,k}\hat{\mathbf{x}}_{k-1|k-1}$
2: Compute $\lambda(\mathbf{x}_k^{(0)})$ using Equation (58)
3: **for** $i = 0, 1, \ldots$ **do**
4:    Compute $\psi(\mathbf{x}_k^{(i)}|\mathcal{Z}_k)$ using Equations (55) and (56)
5:    Set $\mathbf{R}_k^{(i)} = \psi(\mathbf{x}_k^{(i)}|\mathcal{Z}_k)\bar{\mathbf{R}}_k$ {see Equation (26)}
6:    Compute $\mathbf{x}_k^{(i+1)}$ and $\mathbf{P}_k^{(i+1)}$ (or its inverse) using the standard Kalman filter recursions with measurement noise covariance matrix $\mathbf{R}_k = \mathbf{R}_k^{(i)}$ {see, e.g., Equations (29) through (32)}
7:    Compute $\lambda(\mathbf{x}_k^{(i+1)})$ using Equation (58)
8:    **if** $\lambda(\mathbf{x}_k^{(i+1)}) - \lambda(\mathbf{x}_k^{(i)}) < \epsilon$ **then**
9:      break
10:    **end if**
11: **end for**
12: Set $\hat{\mathbf{x}}_{k|k} = \mathbf{x}_k^{(\infty)} = \mathbf{x}_k^{(i+1)}$
13: Set $\mathbf{P}_k^{(\infty)} = \mathbf{P}_k^{(i+1)}$
14: Compute $\phi(\hat{\mathbf{x}}_{k|k}|\mathcal{Z}_k)$ using Equations (55) and (57)
15: Compute $\mathbf{u}_k(\hat{\mathbf{x}}_{k|k})$ using Equation (45)
16: Compute $\hat{\mathbf{P}}_{k|k}$ (or its inverse) using Equation (48) (or Equation (47))

**Figure 3. Pseudo-Code for NVMF.**

### 3.4 CHOOSING THE MIXING DENSITY SHAPE AND SCALE PARAMETERS

The inverse gamma density given by Equation (17) and used here for the measurement noise variance mixing density in the NVMF is a two-parameter density, with shape and scale parameters, $\alpha > 0$ and $\beta > 0$, respectively. Values for these parameters may be derived from the values of two, more intuitive design parameters, namely $r_{\text{out}}$ and $\rho$, where $r_{\text{out}}$ is the maximum expected value of the measurement noise variance, $r_k$, and $\rho$ is the probability $r_k$ exceeds this value; that is, $\rho = \Pr\{r > r_{\text{out}}\}$. In practice, it is desirable to choose $\rho$ to be small so the tails of the NVMF measurement noise distribution are large enough to accomodate the range of outliers expected for the problem of interest.

Once values for the filter design parameters $r_{\text{out}}$ and $\rho$ are chosen, corresponding values for $\alpha$ and $\beta$ may be obtained as follows. First, observe that all four parameters are related by the equalities

$$\rho = 1 - \int_0^{r_{\text{out}}} \mathcal{IG}(r; \alpha, \beta)\, dr = \frac{\gamma(\alpha, \beta/r_{\text{out}})}{\Gamma(\alpha)}, \quad (59)$$

where $\gamma(\cdot, \cdot)$ denotes the lower incomplete gamma function (as defined in Section 6.5.2 of [30]). The first and second equalities follow from the definitions of $\rho$ and the inverse gamma distribution, respectively. The last equality may then be inverted to give $\beta$ as a function of $\alpha$ and the design parameters $r_{\text{out}}$ and $\rho$:

$$\beta = r_{\text{out}}\gamma^{-1}(\alpha, \rho\Gamma(\alpha)), \quad (60)$$

where $\gamma^{-1}(\cdot, \cdot)$ is the inverse lower incomplete gamma function,[§] defined such that

$$\gamma^{-1}(\alpha, \rho\Gamma(\alpha)) = \gamma^{-1}(\alpha, \gamma(\alpha, \beta/r_{\text{out}})) = \beta/r_{\text{out}}. \tag{61}$$

Thus, once the value for the shape parameter, $\alpha$, is obtained (as described below), the corresponding value for the scale parameter, $\beta$, is obtained from Equation (60).

Given values for the design parameters, $r_{\text{out}}$ and $\rho$, a corresponding value for the shape parameter, $\alpha$, may be obtained by imposing a reasonable constraint on the posterior Fisher information matrix derived in Appendix D. In particular, in the absence of outliers and when the measurement noise variance, $r_k$, is known, it is reasonable to equate the complete-data posterior Fisher information matrix under the NVMF model to the posterior Fisher information matrix under the Kalman filter model. The former matrix is given by the first two terms in Equation (143), namely,

$$\mathbf{J}_k = \hat{\mathbf{P}}_{k|k-1}^{-1} + \mathrm{E}_{\mathbf{v}_k}\left[\psi^{-1}(\mathbf{v}_k)\right]\mathbf{H}_k^{\mathrm{T}}\bar{\mathbf{R}}_k^{-1}\mathbf{H}_k, \tag{62}$$

with

$$\psi^{-1}(\mathbf{v}_k) = \frac{M/2 + \alpha}{\frac{1}{2}\mathbf{v}_k^{\mathrm{T}}\bar{\mathbf{R}}_k^{-1}\mathbf{v}_k + \beta}, \tag{63}$$

and the latter matrix is given by Equation (150),

$$\mathbf{J}_k = \hat{\mathbf{P}}_{k|k-1}^{-1} + \frac{1}{r_k}\mathbf{H}_k^{\mathrm{T}}\bar{\mathbf{R}}_k^{-1}\mathbf{H}_k. \tag{64}$$

Using the parameterization of $\beta$ in terms of $\alpha$ and the design parameters, $r_{\text{out}}$ and $\rho$, given by Equation (60), these two expected information matrices are equal if and only if

$$\mathrm{E}_{\mathbf{v}_k}\left[\frac{M/2 + \alpha}{\frac{1}{2}\mathbf{v}_k^{\mathrm{T}}\bar{\mathbf{R}}_k^{-1}\mathbf{v}_k + r_{\text{out}}\,\gamma^{-1}(\alpha, \rho\Gamma(\alpha))}\right] - \frac{1}{r_k} = 0. \tag{65}$$

The value of the shape parameter, $\alpha$, consistent with the given values of $r_{\text{out}}$ and $\rho$ is then obtained as the root of this equation. In practice, an approximation of this root can be found by evaluating the difference on the left-hand side of Equation (65) for a discrete set of possible values for $\alpha$, and choosing the value of $\alpha$ from this set that gives the smallest absolute value of this difference. As discussed in Appendix D, for any given value of $\alpha$, the expectation in Equation (65) can be approximated using random sampling from the $M$-variate $t$ distribution (see Appendix A).

While the procedure described here for choosing the inverse gamma mixing density shape and scale parameters is somewhat tedious, it only needs to be done once when designing the filter for the problem of interest. Potential procedures for estimating these parameters concurrently with the filtering process, as attempted by the approach in [13], are beyond the scope of this research.

---

[§]The inverse incomplete gamma function is implemented in many commonly used numerical computing programs (e.g., as the function `gammaincinv` in MATLAB).

# 4. SIMULATIONS

The NVMF recursions are demonstrated in this section for a simple, two-dimensional tracking problem, both with and without outliers. The peformance of the NVMF is compared to that of the standard Kalman filter, the PDAF, the robust filter derived in [10] (referred to here as the variational inference filter (VIF)), and the Kalman filter for outlier rejection (KFOR) method derived in [4], via three Monte Carlo simulations—each with a different simulated measurement noise distribution. To facilitate fair comparisons between the filters, a concerted effort is made to select the parameters unique to each filter in a consistent manner, as described in Sections 4.3.1 through 4.3.5.

## 4.1 PROBLEM DEFINITION

The two-dimensional tracking problem simulated in this section is the sequential estimation of the position and velocity of an object, moving at roughly constant velocity in the $xy$-plane, from periodic measurements of its position. The state vector in this case is the $L = 4$ element vector consisting of the $xy$ position and velocity of the object at time index, $k$,

$$\mathbf{x}_k = \begin{bmatrix} x_k & y_k & \dot{x}_k & \dot{y}_k \end{bmatrix}^\mathrm{T}. \tag{66}$$

For constant velocity motion, $\mathbf{x}_k$ is related to the state at the previous time index, $k-1$, by the state transition matrix

$$\mathbf{F}_{k-1,k} = \begin{bmatrix} 1 & T_{k-1,k} \\ 0 & 1 \end{bmatrix} \otimes \begin{bmatrix} 1 & 0 \\ 0 & 1 \end{bmatrix}, \tag{67}$$

where $T_{k-1,k}$ is the elapsed time between the two updates. It is assumed the object under track experiences slight peturbations to its intended constant velocity in the form of small, white noise accelerations, leading to a process noise covariance matrix of the form

$$\mathbf{Q}_k = q_k \begin{bmatrix} \frac{1}{3}T_{k-1,k}^3 & \frac{1}{2}T_{k-1,k}^2 \\ \frac{1}{2}T_{k-1,k}^2 & T_{k-1,k} \end{bmatrix} \otimes \begin{bmatrix} 1 & 0 \\ 0 & 1 \end{bmatrix}, \tag{68}$$

where $q_k > 0$ is a small process noise variance (see Section 6.2.2 of [31]). For the simulations, the position measurements are periodic with period, $T$, and the process noise variance is a constant value, $q$. Hence, $T_{k-1,k} = T$ and $q_k = q$ for all values of $k$.

For this problem, the measurement vector, $\mathbf{z}_k$, is the $M = 2$ element vector consisting of measurements of the $xy$ position of the object. Thus, the measurement matrix that maps the state space to the measurement space is simply

$$\mathbf{H}_k = \begin{bmatrix} 1 & 0 & 0 & 0 \\ 0 & 1 & 0 & 0 \end{bmatrix}. \tag{69}$$

## 4.2 SIMULATED MEASUREMENT NOISE DISTRIBUTIONS

For the three simulations, each measurement, $\mathbf{z}_k$, is generated according to the model given by Equation (2) where, depending on the simulation, the measurement noise vector, $\mathbf{v}_k$, is drawn from one of three distributions:

1. Gaussian,
$$p(\mathbf{v}_k) = \mathcal{N}(\mathbf{v}_k; \mathbf{0}, \bar{r}\bar{\mathbf{R}}), \tag{70}$$

where $\bar{r}$ denotes the "regular" measurement noise variance (i.e., when no outliers are present), and $\bar{\mathbf{R}}$ is the $2 \times 2$ identity matrix;

2. Gaussian-uniform (GU) mixture,
$$p(\mathbf{v}_k) = (1 - P_{\text{out}})\mathcal{N}(\mathbf{v}_k; \mathbf{0}, \bar{r}\bar{\mathbf{R}}) + P_{\text{out}}\mathcal{U}(\mathbf{v}_k; -6\sqrt{r_{\text{out}}}, 6\sqrt{r_{\text{out}}}), \tag{71}$$

where $P_{\text{out}}$ is the probability the measurement is an outlier, and $\mathcal{U}(\mathbf{v}_k; -6\sqrt{r_{\text{out}}}, 6\sqrt{r_{\text{out}}})$ denotes the multivariate uniform density with support over the interval $[-6\sqrt{r_{\text{out}}}, 6\sqrt{r_{\text{out}}}]$ for each element of the noise vector, $\mathbf{v}_k$;

3. Multivariate $t$,
$$p(\mathbf{v}_k) = \int_0^\infty \mathcal{N}(\mathbf{v}_k; \mathbf{0}, r\bar{\mathbf{R}})\,\mathcal{IG}(r; \alpha, \beta)\,dr, \tag{72}$$

where $\alpha$ and $\beta$ are the inverse gamma distribution shape and scale parameters, respectively.

The values used for the parameters of these distributions in the simulations are listed in Table 1. The values used for the parameters specific to each filter are also listed in this table; these filter-specific parameters are described in detail in Sections 4.3.1 through 4.3.5.

## 4.3 FILTER-SPECIFIC PARAMETER VALUES

### 4.3.1 Kalman Filter

The standard Kalman filter assumes the measurement noise covariance matrix, $\mathbf{R}_k$, is known. For these simulations $\mathbf{R}_k$ is set to $\bar{r}\bar{\mathbf{R}}$. Consequently, since the measurement noise distribution given by Equation (70) matches that assumed here by the Kalman filter, this filter is the optimal filter (in the MMSE sense) for the first of the three simulations.

### 4.3.2 PDAF

As discussed in Section 1, the PDAF includes an outlier (or "clutter") process model along with the model for the process of interest. A measurement then associates to either the process of interest or the outlier process with certain probabilities. These association probabilities are parameterized by the probability of detecting the object under track at each update, denoted by $P_D$, and a positive scalar, $g$, called the gating threshold, used to define a measurement validation region (or gate). For these simulations the detection probability, $P_D$, is defined in terms of the (small) probability, $\rho$, that the object is not detected, that is,

$$P_D = 1 - \rho. \tag{73}$$

**Table 1. Filter Parameter Values for Simulations.**

| Filter | Parameter | Value |
|---|---|---|
| All | $\mathbf{x}_0$ | $\begin{bmatrix}100 & 100 & 20 & 10\end{bmatrix}^T$ |
| | $T$ | 3 |
| | $q$ | $1 \times 10^{-6}$ |
| | $\bar{\mathbf{R}}$ | $\begin{bmatrix}1 & 0 \\ 0 & 1\end{bmatrix}$ |
| | $\bar{r}$ | 100 |
| | $r_{\text{out}}$ | $100^2$ |
| | $P_{\text{out}}$ | 0.1 |
| KF | $\mathbf{R}_k$ | $\bar{r}\bar{\mathbf{R}}$ |
| PDAF | $\rho$ | 0.01 |
| | $P_D = P_D(\rho)$ | 0.99 |
| | $g = g(r_{\text{out}}, \bar{r})$ | 100 |
| NVMF | $\alpha$ | 0.9987 |
| | $\beta = \beta(\alpha, \rho, r_{\text{out}})$ | 99.84 |
| VIF | $h = h(\alpha, \beta, \bar{r})$ | 9.502 |
| KFOR | $\tau$ | 3 |
| | $w = w(r_{\text{out}})$ | 300 |

The gating threshold is defined here in terms of the ratio of the maximum expected outlier variance and regular measurement noise variance,

$$g^2 = r_{\text{out}}/\bar{r}. \tag{74}$$

The gating threshold is used to determine the probability, $P_G$, that a measurement falls in the validation region. The probabilities $P_D$ and $P_G$ are then used in the calculation of the measurement association probabilities as shown in Section 3 of [5].

### 4.3.3 NVMF

Values for the NVMF measurement noise variance shape and scale parameters, $\alpha$ and $\beta$, are chosen for these simulations using the procedure discussed in Section 3.4. These values are dictated by the values listed in Table 1 for the design parameters, $r_{\text{out}}$ and $\rho$.

*4.3.4 VIF*

In [10], the measurement noise covariance matrix, $\mathbf{R}_k$, is assumed to be inverse Wishart distributed with $M \times M$, positive-definite scale matrix, $h\mathbf{\Psi}$, and degrees of freedom, $h > M - 1$ (see Sections 10.33 and 10.34 in [32] for a description and properties of the inverse Wishart distribution). Based on equation (23) in Section III.C of [10] and the discussion in the last paragraph of that section, the matrix $\mathbf{\Psi}$ is obtained in terms of the regular measurement noise parameters defined here via a limiting argument as $h \to \infty$, resulting in

$$\mathbf{\Psi} = \bar{r}\bar{\mathbf{R}}. \tag{75}$$

For consistency with the parameter values chosen for the NVMF, the parameter $h$ in the VIF is chosen by matching the modes of the distributions of $\mathbf{R}_k$ under the NVMF and VIF measurement noise models. Under the NVMF model, where $\mathbf{R}_k = r_k \bar{\mathbf{R}}$, and $r_k$ is inverse gamma distributed as described in Section 4.3.3, the mode is given by

$$\mathrm{mode}(\mathbf{R}_k) = \frac{\beta}{\alpha + 1} \bar{\mathbf{R}}. \tag{76}$$

Under the VIF model, where $\mathbf{R}_k$ is inverse Wishart distributed as described above, the mode is given by

$$\mathrm{mode}(\mathbf{R}_k) = \frac{h\bar{r}}{h + M + 1} \bar{\mathbf{R}}. \tag{77}$$

Equating Equations (76) and (77) and solving for $h$ yields

$$h = \frac{\left(\frac{\beta}{\alpha+1}\right)/\bar{r}}{1 - \left(\frac{\beta}{\alpha+1}\right)/\bar{r}} (M + 1). \tag{78}$$

Lastly, both the NVMF and VIF, unlike the KF and PDAF, are iterative algorithms; several iterations are required at each update, $k$, for them to converge to their state estimates. For simplicity, in both cases a fixed, conservative number of 25 iterations is chosen over performing convergence tests. However, anecdotal evidence based on these simulations suggests both algorithms largely converge after many fewer iterations.

*4.3.5 KFOR*

Like the PDAF, the KFOR includes a distinct model for measurement outliers. However, unlike the PDAF, the KFOR does not assign association probabilities to each measurement, but rather accepts or rejects each measurement as an outlier based on the outcome of a statistical test, and adjusts the measurement noise covariance matrix accordingly. In particular, the KFOR declares the $m$th component of the $k$th measurement, denoted $\mathbf{z}_{k,m}$, to be an outlier if its normalized residual exceeds a specified threshold, $\tau$, that is,

$$\mathbf{i}_{k,m} = \begin{cases} 1, & \frac{|\tilde{\mathbf{z}}_{k,m}|}{\sqrt{\hat{\mathbf{R}}_{k,m}}} > \tau, \\ 0, & \text{otherwise}, \end{cases} \tag{79}$$

where $\mathbf{i}_{k,m}$ is the $m$th component of the outlier indicator vector, $\mathbf{i}_k$, associated with $\mathbf{z}_k$, and

$$\tilde{\mathbf{z}}_k = \mathbf{z}_k - \hat{\mathbf{z}}_{k|k-1}, \tag{80}$$

$$\hat{\mathbf{R}}_k = \mathbf{H}_k \hat{\mathbf{P}}_{k|k-1} \mathbf{H}_k^{\mathrm{T}} + \mathbf{R}_k, \tag{81}$$

are the measurement residual vector and predicted measurement noise covariance matrix, respectively. Once the outlier indicator vector is computed, and assuming the outlier distribution is uniform, with zero mean and half-width, $w$, the measurement noise covariance matrix used by the KFOR for the $k$th update is

$$\mathbf{R}'_k = \mathbf{R}_k + \frac{1}{3} w^2 \operatorname{diag}(\mathbf{i}_k). \tag{82}$$

Hence, the $k$th update of the KFOR is equivalent to that of the standard Kalman filter if no outliers are detected (i.e., if $\mathbf{i}_k = \mathbf{0}$). For these simulations, the half-width of the uniform distribution for the outlier noise model is taken as a multiple of the maximum expected outlier standard deviation, namely,

$$w = 3\sqrt{r_{\text{out}}}. \tag{83}$$

## 4.4 FILTER INITIALIZATION

For tracking an object moving with near constant velocity, the two-point differencing scheme discussed in Section 5.5.3 of [31] may be used to initialize the filters. In this scheme, the initial conditions, $\hat{\mathbf{x}}_{0|0}$ and $\hat{\mathbf{P}}_{0|0}$, are estimated from measurements obtained (or generated, in the case of the simulations) at time indices $k = -1$ and $k = 0$. Assuming these measurements are not outliers and follow the standard measurement model given by Equations (2) and (4), this scheme yields values for $\hat{\mathbf{x}}_{0|0}$ and $\hat{\mathbf{P}}_{0|0}$ given by

$$\hat{\mathbf{x}}_{0|0} = \begin{bmatrix} \mathbf{z}_0 \\ (\mathbf{z}_0 - \mathbf{z}_{-1})/T_{0,-1} \end{bmatrix} \tag{84}$$

and

$$\hat{\mathbf{P}}_{0|0} = \begin{bmatrix} \mathbf{R}_0 & \mathbf{R}_0/T_{0,-1} \\ \mathbf{R}_0/T_{0,-1} & (\mathbf{R}_0 + \mathbf{R}_{-1})/T_{0,-1}^2 \end{bmatrix}, \tag{85}$$

respectively. As discussed in [31], this initialization scheme guarantees consistency of the initial estimate $\hat{\mathbf{x}}_{0|0}$; that is, it guarantees the estimation error covariance matrix for $\hat{\mathbf{x}}_{0|0}$ is equal to $\hat{\mathbf{P}}_{0|0}$ (see Section 4.5 for more on estimator consistency). For the simulations in this section, the initial conditions simplify to

$$\hat{\mathbf{x}}_{0|0} = \begin{bmatrix} \mathbf{z}_0 \\ (\mathbf{z}_0 - \mathbf{z}_{-1})/T \end{bmatrix} \tag{86}$$

and

$$\hat{\mathbf{P}}_{0|0} = \begin{bmatrix} 1 & 1/T \\ 1/T & 2/T^2 \end{bmatrix} \otimes \bar{r}\bar{\mathbf{R}}. \tag{87}$$

## 4.5 PERFORMANCE STATISTICS

For each simulation, filter error and consistency statistics are calculated for each filter. These statistics (listed below) are based on averages of quadratic forms of the state estimation error for the $k$th update of the $n$th simulation trial (out of $N$ total trials), denoted by

$$\tilde{\mathbf{x}}_{k|k}(n) = \hat{\mathbf{x}}_{k|k}(n) - \mathbf{x}_k, \tag{88}$$

where $\hat{\mathbf{x}}_{k|k}(n)$ is the estimate of the state at time index $k$ given all measurements up through $k$ for the $n$th trial, with estimation error covariance matrix $\hat{\mathbf{P}}_{k|k}(n)$.

### 4.5.1 Filter Error

The error statistic of most interest for filter comparisons is the (sample) mean squared error (MSE):

$$\text{MSE}_{k|k} = \frac{1}{N} \sum_{n=1}^{N} \text{SE}_{k|k}(n), \tag{89}$$

where the squared state estimation error for the $k$th update of the $n$th simulation trial is given by

$$\text{SE}_{k|k}(n) = \tilde{\mathbf{x}}_{k|k}^{\text{T}}(n)\, \tilde{\mathbf{x}}_{k|k}(n). \tag{90}$$

When no outliers are present and the measurement noise statistics match the Kalman filter assumptions, the Kalman filter is the minimum MSE filter and, furthermore, meets the posterior Cramér-Rao lower bound (PCRLB) on estimation error in this case (see Section III.A in [33] and discussion proceeding Equation (150) in Appendix D). Ideally, the MSE performance of the robust filters would approach that of the Kalman filter in this case.

To facilitate comparisons with the Kalman filter and address potential large differences in the dynamic range of MSE over the evolution of each simulation, it is convenient to instead consider the normalized root-MSE (NRMSE), defined as

$$\text{NRMSE}_{k|k} = \sqrt{\text{MSE}_{k|k} \Big/ \text{tr}\, \hat{\mathbf{P}}_{k|k}^{\text{KF}}}, \tag{91}$$

where $\hat{\mathbf{P}}_{k|k}^{\text{KF}}$ denotes the Kalman filter state estimation error covariance matrix, given by the inverse of Equation (150). Hence, in the absence of outliers, one would expect the NRMSE values for the Kalman filter to equal one, on average. Ideally, the NRMSE values for the robust filters examined here would be close to one, with or without outliers present.

### 4.5.2 Filter Consistency

When no outliers are present, and under the statistical assumptions of the standard Kalman filter, the following properties hold:

$$\text{E}(\tilde{\mathbf{x}}_{k|k}(n)) = \mathbf{0}, \tag{92}$$

$$\text{E}(\tilde{\mathbf{x}}_{k|k}(n)\tilde{\mathbf{x}}_{k|k}^{\text{T}}(n)) = \hat{\mathbf{P}}_{k|k}(n). \tag{93}$$

In general and in practice, a state estimator is said to be consistent if its estimation errors have these two properties (see Section 5.4.2 of [31]).

For a given simulation, define the average normalized estimation error squared (ANEES) at the $k$th update as follows:

$$\text{ANEES}_{k|k} = \frac{1}{N} \sum_{n=1}^{N} \tilde{\mathbf{x}}_{k|k}^{\text{T}}(n) \hat{\mathbf{P}}_{k|k}^{-1}(n) \tilde{\mathbf{x}}_{k|k}(n). \tag{94}$$

If the estimator is consistent, then the quadratic form in Equation (94) is chi-square distributed with $L$ degrees of freedom and, by properties of chi-square distributed random variables, $N \cdot \text{ANEES}_{k|k}$ is chi-square distributed with $NL$ degrees of freedom. Hence, a reasonable test for estimator consistency is a test of this simple hypothesis. This is a two-sided test, with the alternative hypothesis being $N \cdot \text{ANEES}_{k|k}$ has degrees of freedom less than or greater than $NL$. For a fixed level of significance, $s$, the acceptance region for this test is the interval

$$\mathcal{A}(1 - s; NL) = [\chi_{NL}^2(s/2), \chi_{NL}^2(1 - s/2)], \tag{95}$$

where $\chi_{NL}^2(\xi)$ denotes the point in the interval $[0, \infty)$ such that the left-tail probability of the chi-square distribution with degrees of freedom $NL$ is $\xi$. Simply put, if the estimator is consistent, then on average $(1 - s)\%$ of the ANEES values $\text{ANEES}_{k|k}$ fall in the interval

$$\mathcal{A}_N(1 - s) = \{a : Na \in \mathcal{A}(1 - s; NL)\} \tag{96}$$

for any given value of the time index $k$.

### 4.5.3  Filter Divergence

A filter can diverge, or exhibit non-decreasing estimation error, if it becomes inconsistent, particularly if it becomes overly optimistic in its self-assessment of estimation error. For these simulations, filter divergence is defined relative to the performance of the Kalman filter under the same conditions and for a time index, $k^*$, after which the filter is assumed to be in steady state. In particular, for any given trial, $n \in \{1, 2, \ldots, N\}$, a robust filter is said to have diverged if its squared error in steady state (i.e., at any time index $k > k^*$) exceeds the upper envelope of the Kalman filter squared error over all trials; that is, the filter is declared diverged if

$$\exists\, k > k^* : \text{SE}_{k|k}(n) > \max_{\ell} \text{SE}_{k|k}^{\text{KF}}(\ell), \tag{97}$$

where $\text{SE}_{k|k}^{\text{KF}}(\ell)$ denotes the squared state estimation error for the $k$th update of the $\ell$th trial of the Kalman filter. The logic behind this choice of filter divergence criterion will become apparent in the next section.

## 4.6  RESULTS

Results for the three simulations corresponding to the three measurement noise distributions listed in Equations (70) through (72) are shown in Figures 4 through 6. Each simulation consists of $N = 1000$ trials, with $K = 600$ updates per trial and one measurement per update. For each of the three simulations, trials for which any of the filters diverge for any update $k > 0$ are discarded. Hence, the effective number of trials for a simulation is $N_{\text{eff}} \leq 1000$. A significance level of $s = 0.95$ is chosen for filter consistency assessment. Thus, if a filter is consistent, 95% of its ANEES values for a given simulation would fall in the interval $\mathcal{A}_{N_{\text{eff}}}(0.05)$, on average. The filters are assumed to have reached steady state by time index $k^* = 150$, or one quarter of the way into the tracking interval.

Figure 4 shows the performance of each filter when the measurement noise statistics match the standard Kalman filter assumptions (i.e., under Gaussian noise). As expected, both the Kalman filter and PDAF meet the PCRLB in this case; that is, the '+' and 'o' markers overlay the dashed line in Figure 4(a) (on average). Likewise, both filters are consistent; that is, the majority of the '+' and 'o' markers fall within the acceptance region (dashed lines) in Figure 4(b). (Because these filters perform nearly identically in this case, and due to the order of plotting, the PDAF markers largely obscure those for the Kalman filter in these plots.) The KFOR also performs well, with NRMSE and ANEES values only slightly larger than those for the Kalman filter and PDAF. The NVMF and VIF both exhibit significantly larger NRMSE values, and the VIF is inconsistent; specifically, the VIF is overly optimistic, i.e., its state estimation error covariance matrices are smaller than they should be, resulting in ANEES values larger than expected. While the NVMF is initially overly optimistic in its state estimation error self-assessment, its ANEES values trend downward toward consistent values after about time index 100; hence, it exhibits consistency in steady state.

Figure 5 shows the performance of each filter when the measurement noise statistics are similar to those assumed by the PDAF (i.e., a GU mixture noise model). As expected, the PDAF exhibits the best performance in this case; it has the lowest NRMSE of all the filters and is consistent throughout the tracking interval. The KFOR also performs well, with slightly higher NRMSE values (after initial large transients), and consistent ANEES values in steady state. While the NVMF exhibits higher NRMSE values than the PDAF and KFOR, it is also consistent in steady state. The VIF exhibits the largest NRMSE values in this case and is also inconsistent throughout the tracking interval. Results for the Kalman filter are not shown in these plots, as it expectedly performs poorly in the presence of outliers, and its NRMSE and ANEES values are beyond the limits of these axes.

Figure 6 shows the performance of each filter when the measurement noise statistics match those assumed by the NVMF (i.e., a multivariate $t$ noise model). As expected, the NVMF exhibits the best performance in this case; it has the lowest NRMSE of all the filters and is consistent in steady state. Interestingly, none of the other filters perform well in this case; each has a similar but higher steady state NRMSE than the NVMF, and none are consistent. In fact, all of the other filters are overly optimistic in their self-assessments of state estimation error. As discussed in Section 4.5, this over-optimism can lead to filter divergence (or "track loss" in the context of this target tracking problem). Filter divergence in these simulations is examined further below. As in Figure 5, results for the Kalman filter are not shown in these plots, as it expectedly performs poorly in the presence of heavy-tailed noise; its NRMSE values are much larger than those of the other filters, and its ANEES values are beyond the axis limits.

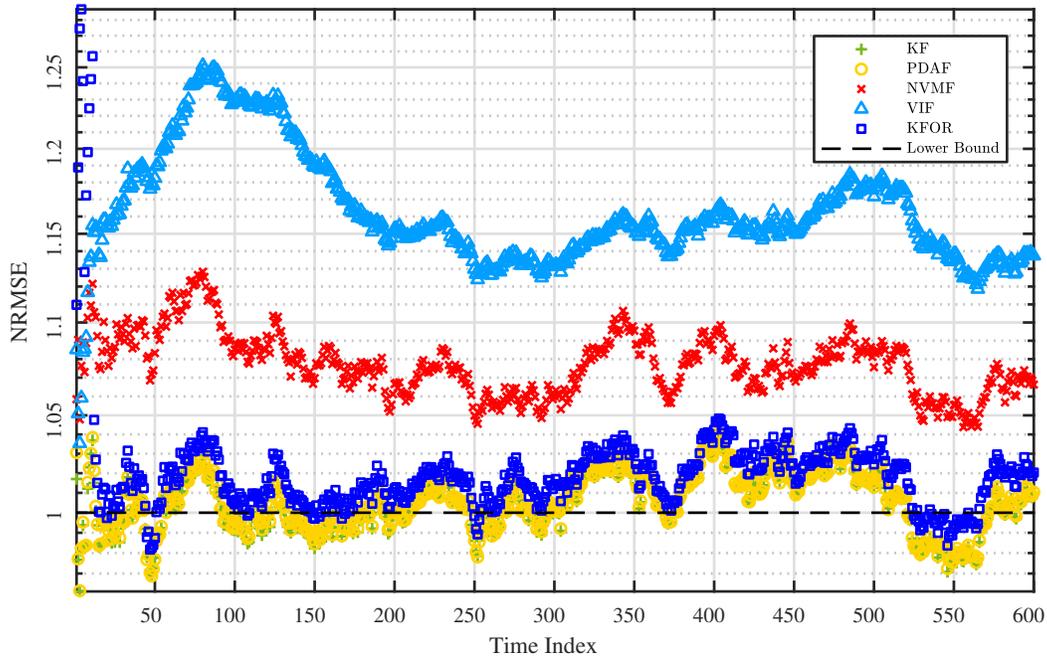

**(a) Estimation Error**

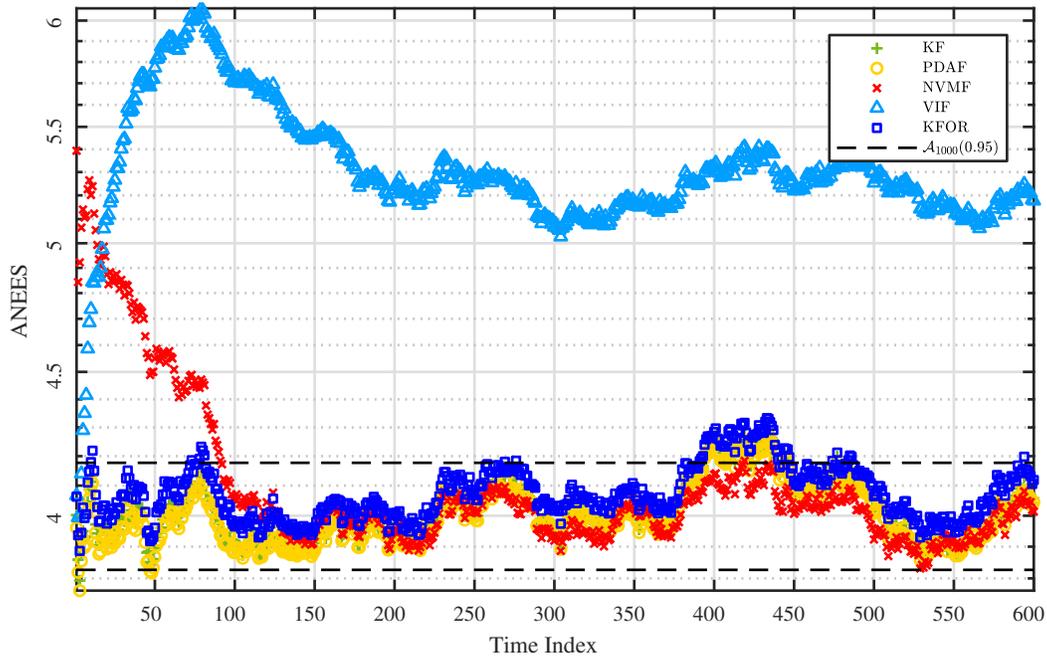

**(b) Consistency**

**Figure 4. Filter Performance for Gaussian Noise.**

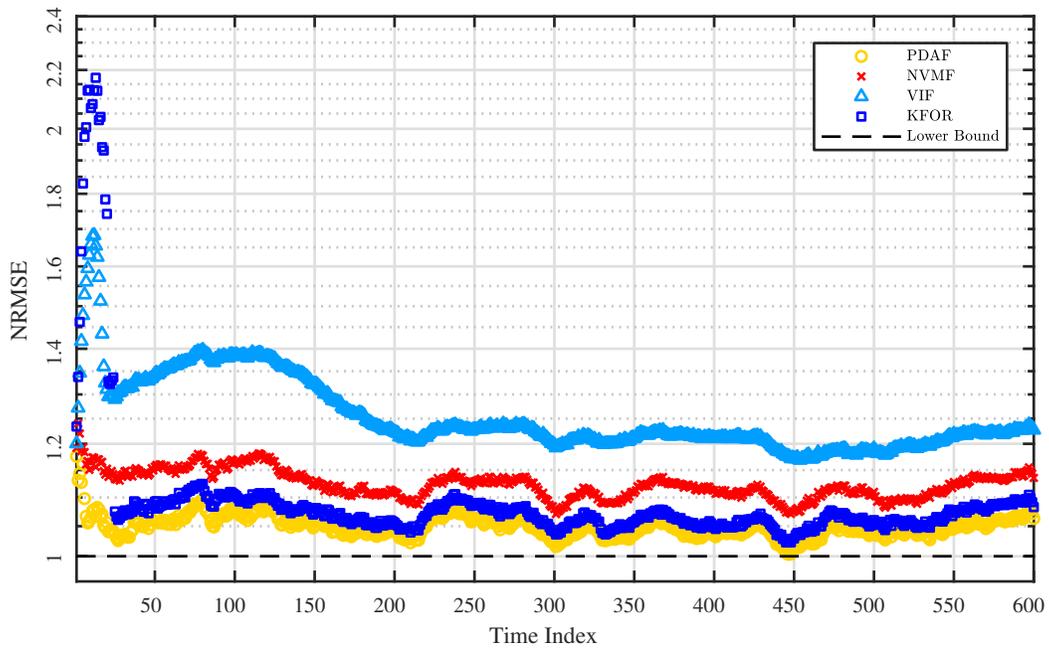

**(a) Estimation Error**

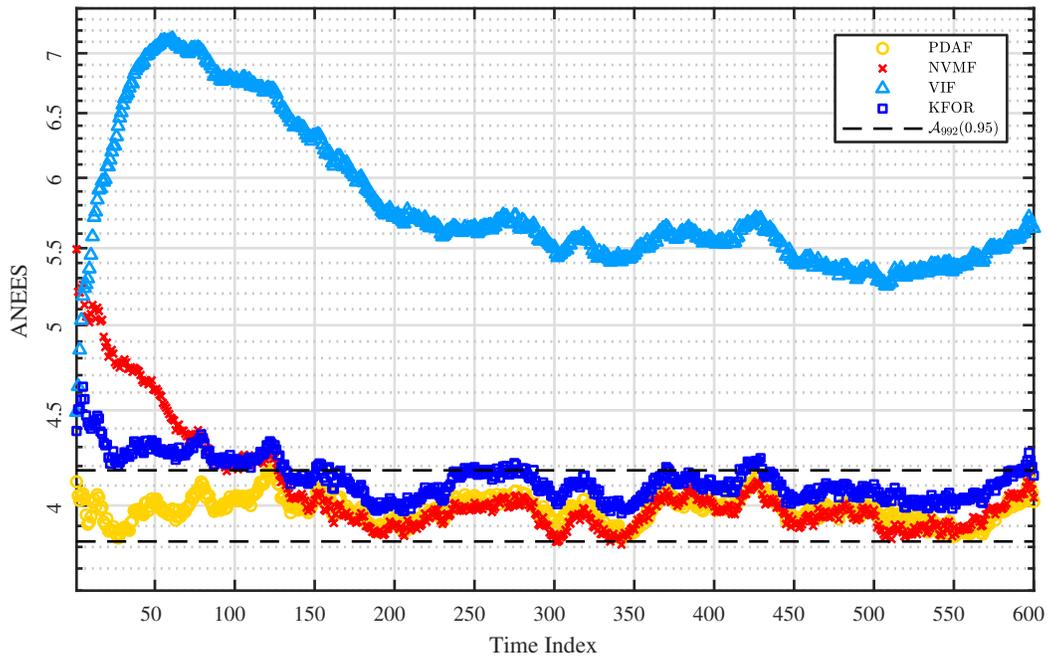

**(b) Consistency**

**Figure 5. Filter Performance for GU Mixture Noise.**

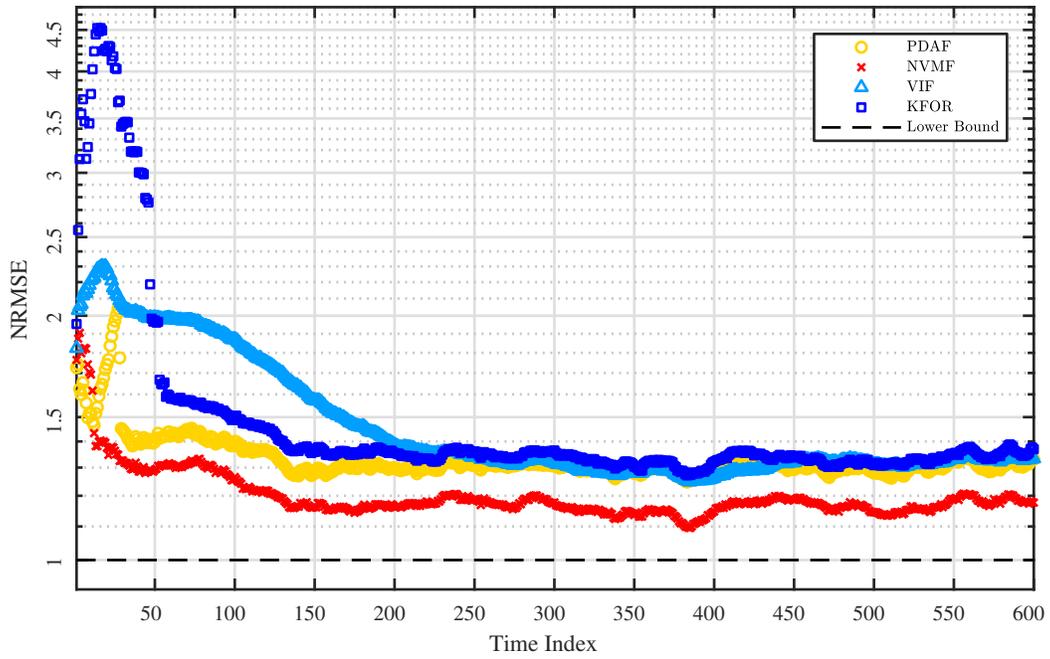

**(a) Estimation Error**

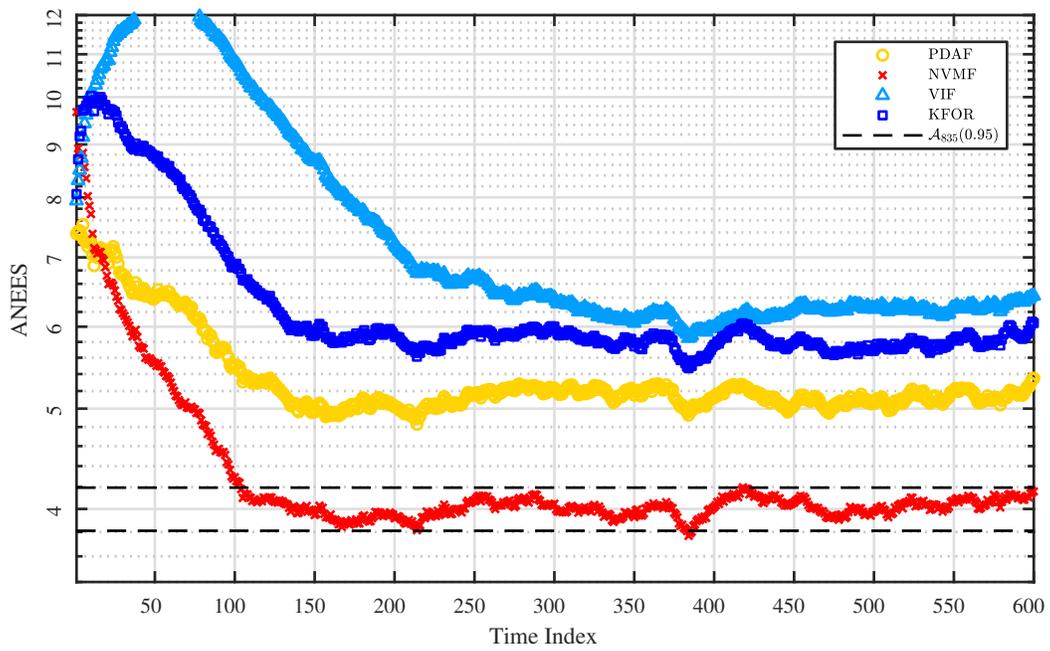

**(b) Consistency**

**Figure 6. Filter Performance for Multivariate $t$ Noise.**

Finally, Figure 7 shows the squared state estimation errors for the Kalman filter, PDAF, and NVMF for each of the 1000 trials (before any are discarded) for the multivariate $t$ measurement noise simulation. The upper envelope of the Kalman filter squared error is shown by the solid green line in each of the last two plots. While both the PDAF and NVMF outperform the Kalman filter in this case, both filters struggle with the two-point differencing scheme for initialization in the presence of heavy-tailed noise. Nevertheless, the NVMF is able to recover and not diverge in any of the 1000 trials. The PDAF, however, diverges in 41 trials. Similar results hold for the VIF and KFOR. The numbers of lost tracks for each filter for each of the three simulations are listed in Table 2.

Table 2. Number of Lost Tracks (out of 1000) for Each Filter.

| Filter | Measurement Noise Model | | |
|---|---|---|---|
| | Gaussian | GU Mixture | Multivariate $t$ |
| KF | - | - | - |
| PDAF | 0 | 0 | 41 |
| NVMF | 0 | 0 | 0 |
| VIF | 0 | 2 | 36 |
| KFOR | 0 | 3 | 28 |

## 4.7 DISCUSSION

It is clear from these results that the standard Kalman filter performs poorly in the presence of outliers while the robust filters (mostly) perform well, and that the robust filters can perform as well (or nearly as well) as the Kalman filter when outliers are not present. The following observations can be made about each of the robust filters:

• The PDAF performs as well as the Kalman filter when no outliers are present. It is the best performer when the actual measurement noise statistics effectively match those assumed by the filter but can perform inconsistently when those statistics are mismatched. The filter can also diverge in the latter case, though infrequently.

• The KFOR performs similarly to—though always worse than—the PDAF.

• While the NVMF exhibits larger errors than the PDAF and the KFOR when no outliers are present and when the outlier statistics follow a uniform distribution, it performs better than both filters when the outliers are multivariate $t$ distributed (unsurprisingly). Furthermore, the NVMF is consistent (in steady state) in all three cases and does not diverge in any of the simulations.

• The VIF is the worst performer; it always exhibits the highest errors and is never consistent. It is possible that this filter's performance could be improved with a different scheme for setting its design parameters than the one proposed in Section 4.3.4, but such an investigation is beyond the scope of this report.

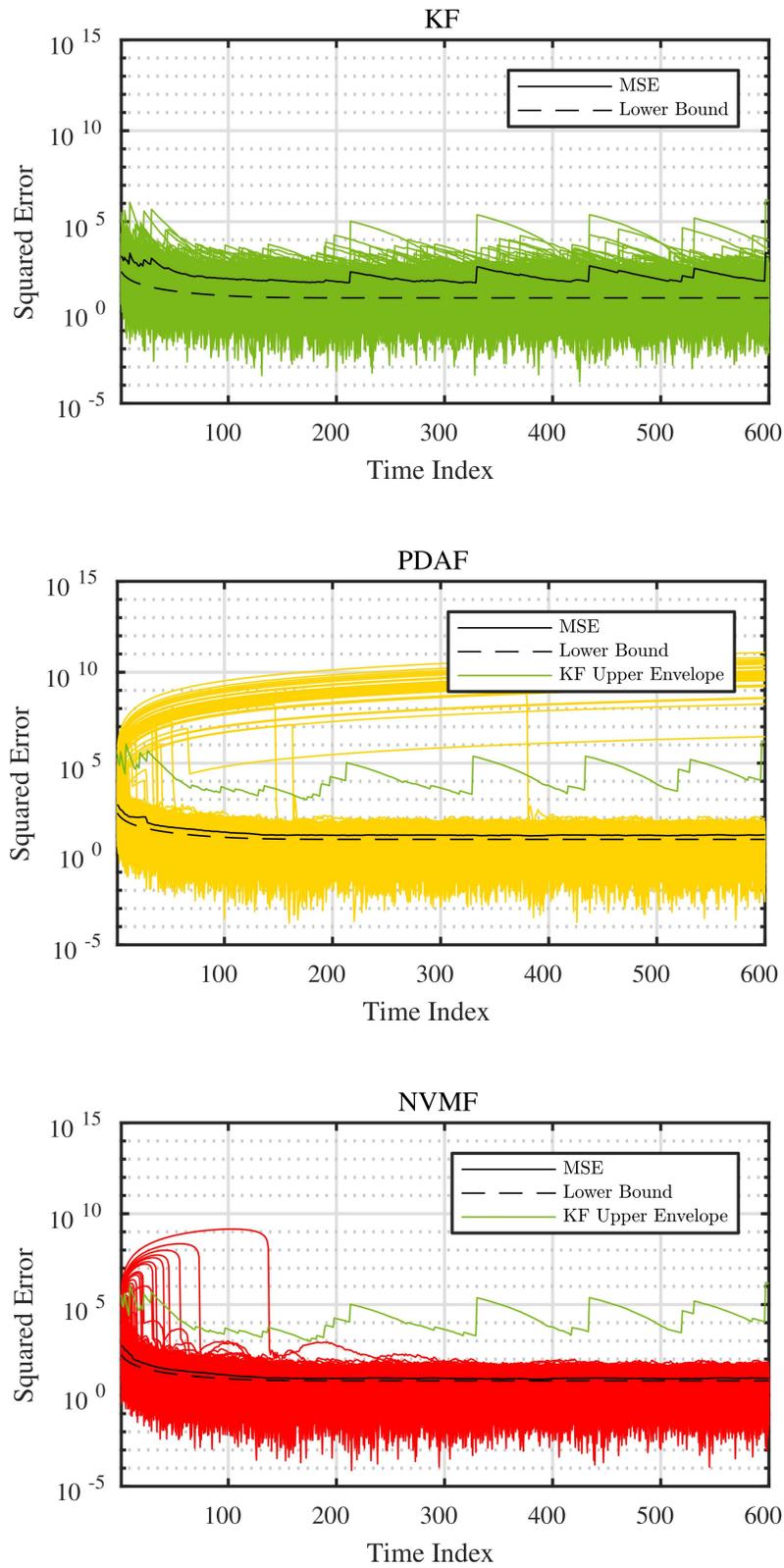

**Figure 7. Track Loss for Multivariate $t$ Noise.**

# 5. SUMMARY AND CONCLUSION

A new filter, similar to the Kalman filter but robust to measurement outliers, is derived in this report. This filter—called the NVMF—uses a normal variance mixture model for the measurement noise distribution, with heavy tails that are better able to accommodate outliers than the comparatively light tails of the Gaussian distribution assumed by the Kalman filter. The EM method and Louis' method are used to derive general recursions for the state estimate and its error covariance matrix, respectively, for the NVMF. When the mixing density for the measurement noise variance is the inverse gamma density, both recursions have simple, closed-form expressions.

The NVMF is compared with the Kalman filter, the PDAF, a similar filter derived using variational inference (referred to here as the VIF), and an outlier detection/adaptation approach (the KFOR) for a simple, two-dimensional tracking problem, both with and without outliers. While the Kalman filter, PDAF, and KFOR provide the best performance when no outliers are present, the NVMF also performs well in terms of consistency in this case.

However, the robust filters clearly outperform the Kalman filter when outliers are present, as expected. The PDAF gives the best performance in terms of MSE when the outliers are uniformly distributed; likewise, the NVMF gives the best MSE performance when the outliers are multivariate $t$ distributed. These results are not surprising—indeed, a robust filter performs best when the statistics of its measurement noise model match the actual noise statistics. However, the PDAF is inconsistent in the multivariate $t$ case, and also exhibits occasional divergence. In contrast, the NVMF is consistent in both the multivariate $t$ and GU noise cases, and never diverges (at least in the simulations performed here).

Thus, it would appear from these simulations that the NVMF is more robust to mismatch in the outlier noise statistics than the PDAF, though this increased robustness comes at the cost of requiring several iterations (i.e., more computations) at each update. However, anecdotal evidence from the simulations conducted here suggest the NVMF converges in relatively few (less than 10) iterations. Moreover, acceleration of EM algorithm convergence is a well-studied problem. Application of acceleration methods to the NVMF is a topic for future research.

Lastly, the NVMF approach to robust filtering extends to the smoothing problem, whereas the PDAF approach does not. The normal variance mixture approach to smoothing is partially explored in [18] but requires further study. In particular, neither the forward nor backward recursions for the estimation error covariance matrix for the smoothing problem are considered in [18]. Development of these recursions, along with comparative performance assessments for the smoothing problem, are subjects of ongoing research.

# APPENDIX A—EQUIVALENCE OF NORMAL VARIANCE MIXTURE AND MULTIVARIATE $t$ DISTRIBUTIONS

Assuming an inverse gamma distribution for the mixing parameter, $r$, in the normal variance mixture PDF given by Equation (5), with shape parameter, $\alpha > 0$, scale parameter, $\beta > 0$, and PDF given by Equation (17), it can be shown that the resulting inverse gamma normal variance mixture PDF is equivalent to the PDF of a central (i.e., zero mean) $M$-variate $t$ distribution as defined in Section 1.1 of [20], with degrees of freedom (or shape parameter), $\nu = 2\alpha$, and correlation matrix $\boldsymbol{\Sigma} = (\beta/\alpha)\bar{\mathbf{R}}$; that is, dropping the time index, $k$,

$$p(\mathbf{v}) = \frac{\Gamma\left(\frac{\nu+M}{2}\right)}{(\pi\nu)^{M/2}|\boldsymbol{\Sigma}|^{1/2}\Gamma\left(\frac{\nu}{2}\right)} \left(1 + \frac{1}{\nu}\mathbf{v}^{\mathrm{T}}\boldsymbol{\Sigma}^{-1}\mathbf{v}\right)^{-\frac{\nu+M}{2}}. \tag{98}$$

This equivalence is obtained by substituting Equation (17) into Equation (5) and manipulating the resulting expression so that the following integral can be used:

$$\int_0^\infty \frac{1}{y^{\eta+1}}\, \mathrm{e}^{-\mu/y}\, dy = \frac{1}{\mu^\eta}\Gamma(\eta), \tag{99}$$

for $\mu, \eta > 0$ (this integral is derived from integral number 4 in Section 3.381 of [34] using a simple change of variables). Substituting $\alpha = \nu/2$ and $\bar{\mathbf{R}} = (\alpha/\beta)\boldsymbol{\Sigma}$ and simplifying the resulting expression yields Equation (98).

In the one-dimensional case, if $\boldsymbol{\Sigma} = 1$, then the PDF given by Equation (98) is the PDF of the univariate Student's $t$ distribution with degrees of freedom, $\nu$, with increasingly heavy tails as $\nu \to 1$. In the $M$-dimensional case, the limiting form of Equation (98) as $\nu \to \infty$ is the PDF of the $M$-variate normal distribution with mean vector zero and covariance matrix $\boldsymbol{\Sigma}$. An extensive treatment of the multivariate $t$ distribution and its applications is provided in [20].

# APPENDIX B—COMPLETE AND INCOMPLETE DATA DENSITIES

The EM method and Louis's method used to derive the recursions for the state estimate and its error covariance matrix in Sections 3.1 and 3.2, respectively, require specification of both the complete data posterior PDF, $p(\mathbf{x}_k|\mathcal{Z}_k, r_k)$, and incomplete data posterior PDF, $p(\mathbf{x}_k|\mathcal{Z}_k)$. These densities are derived in this appendix, under the following assumptions:

- $\mathbf{z}_k$ is independent of $\mathcal{Z}_{k-1}$ when conditioned on $\mathbf{x}_k$;
- $r_k$ and $\mathcal{Z}_{k-1}$ are independent;
- $r_k$ and $\mathbf{x}_k$ are independent;
- $\mathbf{x}_k$ is independent of $\mathcal{Z}_{k-1}$ when conditioned on $\mathbf{x}_{k-1}$.

Expressions are derived first for the general case, and then for the linear Gaussian case.

## B.1 GENERAL CASE

### B.1.1 Complete Data Posterior PDF

The complete data posterior PDF is given, in general, by

$$p(\mathbf{x}_k|\mathcal{Z}_k, r_k) = p(\mathbf{x}_k|\mathbf{z}_k, r_k, \mathcal{Z}_{k-1}), \tag{100}$$

$$= \frac{p(\mathbf{x}_k, \mathbf{z}_k, r_k, \mathcal{Z}_{k-1})}{p(\mathbf{z}_k, r_k, \mathcal{Z}_{k-1})}, \tag{101}$$

$$= \frac{p(\mathbf{z}_k, r_k|\mathbf{x}_k, \mathcal{Z}_{k-1}) \, p(\mathbf{x}_k|\mathcal{Z}_{k-1})}{p(\mathbf{z}_k|r_k, \mathcal{Z}_{k-1}) \, p(r_k|\mathcal{Z}_{k-1})}, \tag{102}$$

where the terms, $p(\mathcal{Z}_{k-1})$, implicit in the numerator and denominator in the last two expression, cancel.

The complete data likelihood function (the first term in the numerator of Equation (102)) is given by

$$p(\mathbf{z}_k, r_k|\mathbf{x}_k, \mathcal{Z}_{k-1}) = p(\mathbf{z}_k|\mathbf{x}_k, r_k, \mathcal{Z}_{k-1}) \, p(r_k|\mathbf{x}_k, \mathcal{Z}_{k-1}), \tag{103}$$

$$= p(\mathbf{z}_k|\mathbf{x}_k, r_k) \, p(r_k). \tag{104}$$

The prior PDF (the second term in the numerator of Equation (102)) is given by

$$p(\mathbf{x}_k|\mathcal{Z}_{k-1}) = \int p(\mathbf{x}_k|\mathbf{x}_{k-1}, \mathcal{Z}_{k-1}) \, p(\mathbf{x}_{k-1}|\mathcal{Z}_{k-1}) \, d\mathbf{x}_{k-1}, \tag{105}$$

$$= \int p(\mathbf{x}_k|\mathbf{x}_{k-1}) \, p(\mathbf{x}_{k-1}|\mathcal{Z}_{k-1}) \, d\mathbf{x}_{k-1}, \tag{106}$$

Finally, the first term in the denominator of Equation (102) can be written as

$$p(\mathbf{z}_k|r_k, \mathcal{Z}_{k-1}) = \int p(\mathbf{z}_k|\mathbf{x}_k, r_k, \mathcal{Z}_{k-1}) \, p(\mathbf{x}_k|r_k, \mathcal{Z}_{k-1}) \, d\mathbf{x}_k, \quad (107)$$

$$= \int p(\mathbf{z}_k|\mathbf{x}_k, r_k) \, p(\mathbf{x}_k|\mathcal{Z}_{k-1}) \, d\mathbf{x}_k, \quad (108)$$

which help to simplify expressions in the linear Gaussian case.

### B.1.2 Incomplete Data Posterior PDF

The incomplete data posterior PDF is obtained from the complete data posterior PDF by marginalizing over the missing measurement noise variance, $r_k$:

$$p(\mathbf{x}_k|\mathcal{Z}_k) = \int_0^\infty p(\mathbf{x}_k|\mathcal{Z}_k, r) \, p(r|\mathcal{Z}_k) \, dr. \quad (109)$$

The conditional, missing data PDF in the integrand is given, in general, by

$$p(r_k|\mathcal{Z}_k) = p(r_k|\mathbf{z}_k, \mathcal{Z}_{k-1}), \quad (110)$$

$$= \frac{p(r_k, \mathbf{z}_k, \mathcal{Z}_{k-1})}{p(\mathbf{z}_k, \mathcal{Z}_{k-1})}, \quad (111)$$

$$= \frac{p(\mathbf{z}_k|r_k, \mathcal{Z}_{k-1}) \, p(r_k|\mathcal{Z}_{k-1})}{p(\mathbf{z}_k|\mathcal{Z}_{k-1})}, \quad (112)$$

$$= \frac{p(\mathbf{z}_k|r_k, \mathcal{Z}_{k-1}) \, p(r_k)}{\int_0^\infty p(\mathbf{z}_k|r, \mathcal{Z}_{k-1}) \, p(r) \, dr}. \quad (113)$$

Substituting Equation (104) into Equation (102), substituting this result and Equation (113) into Equation (109) and simplifying the resulting expression yields

$$p(\mathbf{x}_k|\mathcal{Z}_k) = \frac{p(\mathbf{x}_k|\mathcal{Z}_{k-1}) \int_0^\infty p(\mathbf{z}_k|\mathbf{x}_k, r) \, p(r) \, dr}{\int_0^\infty p(\mathbf{z}_k|r, \mathcal{Z}_{k-1}) \, p(r) \, dr}. \quad (114)$$

## B.2 LINEAR GAUSSIAN CASE

Under the linear Gaussian model,

$$p(\mathbf{x}_{k-1}|\mathcal{Z}_{k-1}) = \mathcal{N}(\mathbf{x}_{k-1}; \hat{\mathbf{x}}_{k-1|k-1}, \hat{\mathbf{P}}_{k-1|k-1}), \quad (115)$$

$$p(\mathbf{x}_k|\mathbf{x}_{k-1}) = \mathcal{N}(\mathbf{x}_k; \mathbf{F}_{k-1,k}\mathbf{x}_{k-1}, \mathbf{Q}_k), \quad (116)$$

$$p(\mathbf{z}_k|\mathbf{x}_k, r_k) = \mathcal{N}(\mathbf{z}_k; \mathbf{H}_k\mathbf{x}_k, r_k\bar{\mathbf{R}}_k). \quad (117)$$

Substituting Equation (117) into Equation (104) yields the complete data likelihood function,

$$p(\mathbf{z}_k, r_k|\mathbf{x}_k, \mathcal{Z}_{k-1}) = \mathcal{N}(\mathbf{z}_k; \mathbf{H}_k\mathbf{x}_k, r_k\bar{\mathbf{R}}_k) \, p(r_k). \quad (118)$$

Substituting Equations (115) and (116) into Equation (106) and evaluating the integral using the Gaussian refactorization lemma presented in [35] (with proof given in [36]) yields

$$p(\mathbf{x}_k|\boldsymbol{\mathcal{Z}}_{k-1}) = \mathcal{N}(\mathbf{x}_k; \hat{\mathbf{x}}_{k|k-1}, \hat{\mathbf{P}}_{k|k-1}), \tag{119}$$

where

$$\hat{\mathbf{x}}_{k|k-1} = \mathbf{F}_{k-1,k}\hat{\mathbf{x}}_{k-1|k-1}, \tag{120}$$

$$\hat{\mathbf{P}}_{k|k-1} = \mathbf{Q}_k + \mathbf{F}_{k-1,k}\hat{\mathbf{P}}_{k-1|k-1}\mathbf{F}_{k-1|k}^\mathrm{T}. \tag{121}$$

Substituting Equations (117) and (119) into Equation (108) and evaluating the integral, again using the Gaussian refactorization lemma, yields

$$p(\mathbf{z}_k|r_k, \boldsymbol{\mathcal{Z}}_{k-1}) = \mathcal{N}(\mathbf{z}_k; \mathbf{H}_k\hat{\mathbf{x}}_{k|k-1}, r_k\bar{\mathbf{R}}_k + \mathbf{H}_k\hat{\mathbf{P}}_{k|k-1}\mathbf{H}_k^\mathrm{T}). \tag{122}$$

Finally, substituting Equations (118), (119), and (122) into Equation (102) and simplifying the resulting expression yields the following for the complete data posterior PDF for the linear Gaussian case:

$$p(\mathbf{x}_k|\boldsymbol{\mathcal{Z}}_k, r_k) = \frac{\mathcal{N}(\mathbf{x}_k; \hat{\mathbf{x}}_{k|k-1}, \hat{\mathbf{P}}_{k|k-1})\,\mathcal{N}(\mathbf{z}_k; \mathbf{H}_k\mathbf{x}_k, r_k\bar{\mathbf{R}}_k)}{\mathcal{N}(\mathbf{z}_k; \mathbf{H}_k\hat{\mathbf{x}}_{k|k-1}, r_k\bar{\mathbf{R}}_k + \mathbf{H}_k\hat{\mathbf{P}}_{k|k-1}\mathbf{H}_k^\mathrm{T})}. \tag{123}$$

Likewise, substituting Equations (117), (119), and (122) into Equation (114) yields the following for the incomplete data posterior PDF for the linear Gaussian case:

$$p(\mathbf{x}_k|\boldsymbol{\mathcal{Z}}_k) = \frac{\mathcal{N}(\mathbf{x}_k; \hat{\mathbf{x}}_{k|k-1}, \hat{\mathbf{P}}_{k|k-1}) \int_0^\infty \mathcal{N}(\mathbf{z}_k; \mathbf{H}_k\mathbf{x}_k, r\bar{\mathbf{R}}_k)\,p(r)\,dr}{\int_0^\infty \mathcal{N}(\mathbf{z}_k; \mathbf{H}_k\hat{\mathbf{x}}_{k|k-1}, r\bar{\mathbf{R}}_k + \mathbf{H}_k\hat{\mathbf{P}}_{k|k-1}\mathbf{H}_k^\mathrm{T})\,p(r)\,dr}. \tag{124}$$

It is worth noting that the complete data posterior PDF given by Equation (123) is the posterior PDF for the standard Kalman filter with measurement noise covariance matrix $\mathbf{R}_k = r_k\bar{\mathbf{R}}_k$. Given $r_k$, the MAP estimate (or posterior mode) of $\mathbf{x}_k$ may be found from this PDF in at least two ways. The easiest way is to apply the Gaussian refactorization lemma to the numerator of Equation (123) to obtain

$$\mathcal{N}(\mathbf{z}_k; \mathbf{H}_k\mathbf{x}_k, r_k\bar{\mathbf{R}}_k)\,\mathcal{N}(\mathbf{x}_k; \hat{\mathbf{x}}_{k|k-1}, \hat{\mathbf{P}}_{k|k-1})$$
$$= \mathcal{N}(\mathbf{z}_k; \mathbf{H}_k\hat{\mathbf{x}}_{k|k-1}, r_k\bar{\mathbf{R}}_k + \mathbf{H}_k\hat{\mathbf{P}}_{k|k-1}\mathbf{H}_k^\mathrm{T})\,\mathcal{N}(\mathbf{x}_k; \boldsymbol{\lambda}_k, \boldsymbol{\Lambda}_k), \tag{125}$$

with

$$\boldsymbol{\lambda}_k = \hat{\mathbf{x}}_{k|k-1} + \mathbf{G}_k\left(\mathbf{z}_k - \mathbf{H}_k\hat{\mathbf{x}}_{k|k-1}\right), \tag{126}$$

$$\boldsymbol{\Lambda}_k = (\mathbf{I} - \mathbf{G}_k\mathbf{H}_k)\,\hat{\mathbf{P}}_{k|k-1}, \tag{127}$$

$$\mathbf{G}_k = \hat{\mathbf{P}}_{k|k-1}\mathbf{H}_k^\mathrm{T}\left(r_k\bar{\mathbf{R}}_k + \mathbf{H}_k\hat{\mathbf{P}}_{k|k-1}\mathbf{H}_k^\mathrm{T}\right)^{-1}. \tag{128}$$

Substituting Equation (125) into Equation (123) and canceling terms yields

$$p(\mathbf{x}_k|\boldsymbol{\mathcal{Z}}_k, r_k) = \mathcal{N}(\mathbf{x}_k; \boldsymbol{\lambda}_k, \boldsymbol{\Lambda}_k). \tag{129}$$

Thus, $\hat{\mathbf{x}}_{k|k} = \boldsymbol{\lambda}_k$ and $\hat{\mathbf{P}}_{k|k} = \boldsymbol{\Lambda}_k$, with $\mathbf{G}_k$ the Kalman gain matrix.

Alternatively, the MAP estimate of $\mathbf{x}_k$ given $r_k$ may be obtained in the usual way by finding the value of $\mathbf{x}_k$ that maximizes the logarithm of $p(\mathbf{x}_k|\mathcal{Z}_k, r_k)$ which, dropping terms not dependent on $\mathbf{x}_k$, is given by

$$\log p(\mathbf{x}_k|\mathcal{Z}_k, r_k) = -\frac{1}{2} \left(\mathbf{z}_k - \mathbf{H}_k \mathbf{x}_k\right)^\mathrm{T} \left(r_k \bar{\mathbf{R}}_k\right)^{-1} \left(\mathbf{z}_k - \mathbf{H}_k \mathbf{x}_k\right)$$
$$-\frac{1}{2} \left(\mathbf{x}_k - \hat{\mathbf{x}}_{k|k-1}\right)^\mathrm{T} \hat{\mathbf{P}}_{k|k-1}^{-1} \left(\mathbf{x}_k - \hat{\mathbf{x}}_{k|k-1}\right). \quad (130)$$

Taking the derivative of this equation with respect to $\mathbf{x}_k$, setting the result equal to zero, and solving for $\hat{\mathbf{x}}_{k|k}$ yields the result given by Equation (126).

# APPENDIX C—GRADIENT AND CURVATURE OF COMPLETE DATA LOG LIKELIHOOD FUNCTION AND LOG PRIOR PDF

For the complete data log likelihood function, define its gradient vector, $\mathbf{S}(\mathbf{z}_k, r_k|\mathbf{x}_k)$, and curvature matrix, $\mathbf{B}(\mathbf{z}_k, r_k|\mathbf{x}_k)$, as the first derivative and negative second derivative, respectively:

$$\mathbf{S}(\mathbf{z}_k, r_k|\mathbf{x}_k) = \frac{\partial}{\partial \mathbf{x}_k} \log p(\mathbf{z}_k, r_k|\mathbf{x}_k, \mathcal{Z}_{k-1}), \tag{131}$$

$$= \frac{\partial}{\partial \mathbf{x}_k} \log p(\mathbf{z}_k|\mathbf{x}_k, r_k), \tag{132}$$

$$\mathbf{B}(\mathbf{z}_k, r_k|\mathbf{x}_k) = -\frac{\partial}{\partial \mathbf{x}_k} \mathbf{S}^\mathrm{T}(\mathbf{z}_k, r_k|\mathbf{x}_k). \tag{133}$$

Likewise, define the gradient vector and curvature matrix for the log prior PDF, denoted by $\mathbf{S}(\mathbf{x}_k|\mathcal{Z}_{k-1})$ and $\mathbf{B}(\mathbf{x}_k|\mathcal{Z}_{k-1})$, respectively, as

$$\mathbf{S}(\mathbf{x}_k|\mathcal{Z}_{k-1}) = \frac{\partial}{\partial \mathbf{x}_k} \log p(\mathbf{x}_k|\mathcal{Z}_{k-1}), \tag{134}$$

$$\mathbf{B}(\mathbf{x}_k|\mathcal{Z}_{k-1}) = -\frac{\partial}{\partial \mathbf{x}_k} \mathbf{S}^\mathrm{T}(\mathbf{x}_k|\mathcal{Z}_{k-1}). \tag{135}$$

Then for the linear Gaussian case, from Equation (118),

$$\mathbf{S}(\mathbf{z}_k, r_k|\mathbf{x}_k) = -\mathbf{H}_k^\mathrm{T} \left(r_k \bar{\mathbf{R}}_k\right)^{-1} \left(\mathbf{H}_k \mathbf{x}_k - \mathbf{z}_k\right), \tag{136}$$

$$\mathbf{B}(\mathbf{z}_k, r_k|\mathbf{x}_k) = \mathbf{H}_k^\mathrm{T} \left(r_k \bar{\mathbf{R}}_k\right)^{-1} \mathbf{H}_k, \tag{137}$$

and, from Equation (119),

$$\mathbf{S}(\mathbf{x}_k|\mathcal{Z}_{k-1}) = -\hat{\mathbf{P}}_{k|k-1}^{-1} \left(\mathbf{x}_k - \hat{\mathbf{x}}_{k|k-1}\right), \tag{138}$$

$$\mathbf{B}(\mathbf{x}_k|\mathcal{Z}_{k-1}) = \hat{\mathbf{P}}_{k|k-1}^{-1}. \tag{139}$$

# APPENDIX D—POSTERIOR FISHER INFORMATION MATRIX

Following the development in Section I of [33], the posterior expected (Fisher) information matrix for the $k$th update, denoted $\mathbf{J}_k$, is given by the expected value of the observed information matrix, $\mathbf{J}(\mathbf{x}_k|\mathcal{Z}_k)$, with respect to both $\mathbf{x}_k$ and $\mathcal{Z}_k$,

$$\mathbf{J}_k = \mathrm{E}_{\mathbf{x}_k,\mathcal{Z}_k}[\mathbf{J}(\mathbf{x}_k|\mathcal{Z}_k)]. \tag{140}$$

Using Equation (39), this matrix is written as the sum of two expectations,

$$\mathbf{J}_k = \mathrm{E}_{\mathbf{x}_k,\mathcal{Z}_k}[\mathbf{B}(\mathbf{x}_k|\mathcal{Z}_{k-1})] + \mathrm{E}_{\mathbf{x}_k,\mathcal{Z}_k}[\mathbf{B}(\mathbf{z}_k|\mathbf{x}_k)], \tag{141}$$

where the matrices inside these expectations are the curvature matrices of the log prior PDF and log likelihood function for $\mathbf{x}_k$, respectively. Substituting Equations (40) and (43) into Equation (141) and simplifying the resulting expression using the linearity properties of the expectation operator yields

$$\mathbf{J}_k = \hat{\mathbf{P}}_{k|k-1}^{-1} + \mathrm{E}_{\mathbf{x}_k,\mathcal{Z}_k}\left[\psi^{-1}(\mathbf{x}_k|\mathcal{Z}_k)\right]\mathbf{H}_k^\mathrm{T}\bar{\mathbf{R}}_k^{-1}\mathbf{H}_k \\ - \mathbf{H}_k^\mathrm{T}\bar{\mathbf{R}}_k^{-1}\mathrm{E}_{\mathbf{x}_k,\mathcal{Z}_k}\left[\phi^{-2}(\mathbf{x}_k|\mathcal{Z}_k)\left(\mathbf{H}_k\mathbf{x}_k - \mathbf{z}_k\right)\right. \\ \left. \times (\mathbf{H}_k\mathbf{x}_k - \mathbf{z}_k)^\mathrm{T}\right]\bar{\mathbf{R}}_k^{-1}\mathbf{H}_k, \tag{142}$$

where the functions $\psi^{-1}$ and $\phi^{-1}$ are given by Equations (27) and (44), respectively. By the missing information principle [26], the first two terms in Equation (142) are interpreted as the Fisher information associated with the complete data, while the third term is the Fisher information associated with the missing data.

The posterior Fisher information matrix given by Equation (142) is the general form for the normal variance mixture model for measurement noise given by Equation (5). For choice of the inverse gamma density, Equation (17), for the mixing density, $p(r)$, Equation (142) simplifies by using the change of variables, $\mathbf{v}_k = \mathbf{z}_k - \mathbf{H}_k\mathbf{x}_k$, and Equations (56) and (57) for the functions $\psi^{-1}$ and $\phi^{-1}$. In particular, the posterior Fisher information matrix becomes

$$\mathbf{J}_k = \hat{\mathbf{P}}_{k|k-1}^{-1} + \mathrm{E}_{\mathbf{v}_k}\left[\psi^{-1}(\mathbf{v}_k)\right]\mathbf{H}_k^\mathrm{T}\bar{\mathbf{R}}_k^{-1}\mathbf{H}_k - \mathbf{H}_k^\mathrm{T}\bar{\mathbf{R}}_k^{-1}\mathrm{E}_{\mathbf{v}_k}\left[\phi^{-2}(\mathbf{v}_k)\,\mathbf{v}_k\mathbf{v}_k^\mathrm{T}\right]\bar{\mathbf{R}}_k^{-1}\mathbf{H}_k, \tag{143}$$

where

$$\psi^{-1}(\mathbf{v}_k) = \frac{M/2 + \alpha}{\frac{1}{2}\mathbf{v}_k^\mathrm{T}\bar{\mathbf{R}}_k^{-1}\mathbf{v}_k + \beta}, \tag{144}$$

$$\phi^{-1}(\mathbf{v}_k) = \frac{\sqrt{M/2 + \alpha}}{\frac{1}{2}\mathbf{v}_k^\mathrm{T}\bar{\mathbf{R}}_k^{-1}\mathbf{v}_k + \beta}. \tag{145}$$

While the expectations in Equation (143) cannot be evaluated analytically, they can be approximated using random sampling from the $M$-variate $t$ distribution (see Appendix A).

Incidentally, for choice of the Dirac delta function with support at $r_k$ for the mixing density, that is, for
$$p(r) = \delta(r - r_k), \tag{146}$$
the missing data conditional PDF, Equation (51), becomes
$$p(r|\mathbf{x}_k, \boldsymbol{\mathcal{Z}}_k) = \frac{\mathcal{N}(\mathbf{z}_k; \mathbf{H}_k\mathbf{x}_k, r\bar{\mathbf{R}}_k)\delta(r - r_k)}{\mathcal{N}(\mathbf{z}_k; \mathbf{H}_k\mathbf{x}_k, r_k\bar{\mathbf{R}}_k)}. \tag{147}$$

Substituting this conditional PDF for $r$ into Equations (27) and (44) and evaluating the resulting integrals yields
$$\psi^{-1}(\mathbf{x}_k|\boldsymbol{\mathcal{Z}}_k) = \frac{1}{r_k}, \tag{148}$$
$$\phi^{-1}(\mathbf{x}_k|\boldsymbol{\mathcal{Z}}_k) = 0. \tag{149}$$

Substituting these forms for the functions $\psi^{-1}$ and $\phi^{-1}$ into Equation (142) and simplifying the resulting expression gives
$$\mathbf{J}_k = \hat{\mathbf{P}}_{k|k-1}^{-1} + \frac{1}{r_k}\mathbf{H}_k^\mathrm{T}\bar{\mathbf{R}}_k^{-1}\mathbf{H}_k, \tag{150}$$
which is the posterior Fisher information matrix associated with the standard, linear Gaussian Kalman filter model with measurement noise covariance matrix $\mathbf{R}_k = r_k\bar{\mathbf{R}}_k$. The inverse of this matrix in the covariance matrix associated with the posterior Cramér-Rao lower bound on estimation error for this model (see [33], Section I).

# REFERENCES

1. Rudolph Emile Kalman, "A New Approach to Linear Filtering and Prediction Problems," *Transactions of the ASME–Journal of Basic Engineering*, vol. 82 (Series D), 1960, pp. 35–45.

2. Andrew H. Jazwinski, *Stochastic Processes and Filtering Theory*, Academic Press, 1970.

3. R. L. Stratonovich, "Conditional Markov Processes," *Theory of Probability & Its Applications*, vol. 5, no. 2, 1960, pp. 156–178.

4. E. Navon and B. Z. Bobrovsky, "An Efficient Outlier Rejection Technique for Kalman Filters," *Signal Processing*, vol. 188, 2021.

5. Yaakov Bar-Shalom and Edison Tse, "Tracking in a Cluttered Environment With Probabilistic Data Association," *Automatica*, vol. 11, 1975, pp. 451–460.

6. H. W. Sorenson and D. L. Alspach, "Recursive Bayesian Estimation Using Gaussian Sums," *Automatica*, vol. 7, no. 4, 1971, pp. 465–479.

7. M. Brunot, "A Gaussian Uniform Mixture Model for Robust Kalman Filtering," *IEEE Transactions on Aerospace and Electronic Systems*, vol. 56, no. 4, August 2020, pp. 2656–2665.

8. R. J. Meinhold and N. D. Singpurwalla, "Robustification of Kalman Filter Models," *Journal of the American Statistical Association*, vol. 84, no. 406, 1989, pp. 479–486.

9. G. Agamennoni, J. I. Nieto, and E. M. Nebot, "An Outlier-Robust Kalman Filter," in *IEEE International Conference on Robotics and Automation*, Shanghai, China, 2011.

10. Gabriel Agamennoni, Juan I. Nieto, and Eduardo M. Nebot, "Approximate Inference in State-Space Models With Heavy-Tailed Noise," *IEEE Transactions on Signal Processing*, vol. 60, no. 10, October 2012, pp. 5024–5037.

11. M. Roth, E. Ozkan, and F. Gustafsson, "A Student's $t$ Filter for Heavy Tailed Process and Measurement Noise," in *Proceedings of the International Conference on Acoustics, Speech, and Signal Processing (ICASSP)*, Vancouver, Canada, 2013.

12. M. Roth, *Kalman Filters for Nonlinear Systems and Heavy-Tailed Noise*, Ph.D. thesis, Linköping University, Linköping, Sweden, 2013.

13. Y. Huang, Y. Zhang, P. Shi, Z. Wu, J. Qian, and J. A. Chambers, "Robust Kalman Filters Based on Gaussian Scale Mixture Distributions With Application to Target Tracking," *IEEE Transactions on Systems, Man, and Cybernetics*, vol. 49, no. 10, October 2019, pp. 2082–2096.

14. O. Barndorff-Nielsen, J. Kent, and M. Sørensen, "Normal Variance-Mean Mixtures and $z$ Distributions," *International Statistical Review*, vol. 50, no. 2, August 1982.

15. Mike West, "Outlier Models and Prior Distributions in Bayesian Linear Regression," *Journal of the Royal Statistical Society. Series B (Methodological)*, vol. 46, no. 3, 1984, pp. 431–439.

16. A. P. Dempster, N. M. Laird, and D. B. Rubin, "Maximum Likelihood from Incomplete Data via the EM Algorithm," *Journal of the Royal Statistical Society. Series B (Methodological)*, vol. 39, no. 1, 1977, pp. 1–38.

17. Thomas A. Louis, "Finding the Observed Information Matrix when Using the EM Algorithm," *Journal of the Royal Statistical Society. Series B (Methodological)*, vol. 44, no. 2, 1982, pp. 226–233.

18. Michael J. Walsh and Marcus L. Graham, "Robust Approaches to Long-Term Maneuvering-Target Tracking," NUWC-NPT Technical Report 12,312, Naval Undersea Warfare Center Division, Newport, Rhode Island, 11 September 2019.

19. David M. Blei, Alp Kucukelbir, and Jon D. McAuliffe, "Variational Inference: A Review for Statisticians," *Journal of the American Statistical Association*, vol. 112, no. 518, June 2017, pp. 859–877.

20. Samuel Kotz and Saralees Nadarajah, *Multivariate t Distributions and Their Applications*, Cambridge University Press, 2004.

21. C. F. Jeff Wu, "On the Convergence Properties of the EM Algorithm," *The Annals of Statistics*, vol. 11, no. 1, March 1983, pp. 95–103.

22. Naonori Ueda and Ryohei Nakano, "Deterministic annealing EM Algorithm," *Neural Networks*, vol. 11, no. 2, March 1998, pp. 271–282.

23. Xiao-Li Meng and Donald B. Rubin, "Using EM to Obtain Aysmptotic Variance-Covariance Matrices: The SEM Algorithm," *Journal of the American Statistical Association*, vol. 86, no. 416, December 1991, pp. 899–909.

24. Andrew Gelman, John B. Carlin, Hal S. Stern, David B. Dunson, Aki Vehtari, and Donald B. Rubin, *Bayesian Data Analysis*, `www.stat.columbia.edu/~gelman/book/BDA3.pdf` (accessed 23 September 2024), 2020.

25. Bradley Efron and David V. Hinkley, "Assessing the Accuracy of the Maximum Likelihood Estimator: Observed Versus Expected Fisher Information," *Biometrika*, vol. 65, no. 3, December 1978, pp. 457–482.

26. Terence Orchard and Max A. Woodbury, "A Missing Information Principle: Theory and Applications," in *Proceedings of the Sixth Berkeley Symposium on Mathematical Statistics and Probability*, Lucien M. Le Cam, Jerzy Neyman, and Elizabeth L. Scott, Eds., January 1972, vol. 1, pp. 697–715.

27. Gerald J. Bierman, *Factorization Methods for Discrete Sequential Estimation*, Dover, 2006.

28. Phillip L. Ainsleigh, "Parameter Estimation in Dynamical Models for Application in Signal Classification," NUWC-NPT Technical Memorandum 00-014, Naval Undersea Warfare Center Division, Newport, RI, 6 February 2000.

29. Å. Björck, H. Park, and L. Eldén, "Accurate Downdating of Least Squares Solutions," *SIAM Journal on Matrix Analysis and Applications*, vol. 15, no. 2, 1994, pp. 549–568.

30. Milton Abramowitz and Irene A. Stegun, Eds., *Handbook of Mathematical Functions*, Dover, 1972.

31. Yaakov Bar-Shalom and Xiao-Rong Li, *Estimation and Tracking: Principles, Techniques, and Software*, Artech House, 1993.

32. Anthony O'Hagan, *Kendall's Advanced Theory of Statistics*, vol. 2B: Bayesian Inference, Oxford University Press, 1994.

33. Petr Tichavský, Carlos H. Muravchik, and Arye Nehorai, "Posterior Cramér-Rao Bounds for Discrete-Time Nonlinear Filtering," *IEEE Transactions on Signal Processing*, vol. 46, no. 5, May 1998, pp. 1386–1396.

34. I. S. Gradshteyn and I. M. Ryzhik, *Table of Integrals, Series, and Products*, 4 edition, Academic Press, 1980.

35. Phillip L. Ainsleigh, Nasser Kehtarnavaz, and Roy L. Streit, "Hidden Gauss-Markov Models for Signal Classification," *IEEE Transactions on Signal Processing*, vol. 50, no. 6, June 2002, pp. 1355–1367.

36. Phillip L. Ainsleigh, "Theory of Continuous-State Hidden Markov Models and Hidden Gauss-Markov Models," NUWC-NPT Technical Report 11,274, Naval Undersea Warfare Center Division, Newport, RI, 20 March 2001.